\documentclass[11pt,twoside]{article}
\usepackage[pdftex]{graphicx}
\usepackage{amsmath}
\usepackage{amssymb}
\usepackage{array,dcolumn}
\usepackage{textgreek}
\usepackage[utf8]{inputenc}
\usepackage[T1]{fontenc}
\usepackage{mathptmx}
\usepackage{amsfonts,wrapfig,booktabs,makecell}
\usepackage[caption=false]{subfig}
\usepackage{hyperref,url}
\usepackage{csquotes}
\begin{document}
\newcommand{\pst}{\hspace*{1.5em}}

\newcommand{\rigmark}{\em Journal of Russian Laser Research}
\newcommand{\lemark}{\em Volume XX, Number X, 20XX}

\newcommand{\be}{\begin{equation}}
\newcommand{\ee}{\end{equation}}
\newcommand{\bm}{\boldmath}
\newcommand{\ds}{\displaystyle}
\newcommand{\bea}{\begin{eqnarray}}
\newcommand{\eea}{\end{eqnarray}}
\newcommand{\ba}{\begin{array}}
\newcommand{\ea}{\end{array}}
\newcommand{\arcsinh}{\mathop{\rm arcsinh}\nolimits}
\newcommand{\arctanh}{\mathop{\rm arctanh}\nolimits}
\newcommand{\bc}{\begin{center}}
\newcommand{\ec}{\end{center}}

\thispagestyle{plain}

\label{sh}


\begin{center} {\Large \bf
\begin{tabular}{c}
A THEORY OF REFLECTIVE X-RAY MULTILAYER STRUCTURES  
\\[-1mm]
WITH GRADED PERIOD AND ITS APPLICATIONS
\end{tabular}
 } \end{center}

\bigskip

\bigskip

\begin{center} {\bf
R.M. Feshchenko$^{1}$
}\end{center}

\medskip

\begin{center}
{\it
$^1$P.N. Lebedev Physical Institute, Russian Academy of Sciences, 53 Leninski Prospekt\\
Moscow, 119991, Russia
}
\smallskip

$^*$Corresponding author e-mail: rusl@sci.lebedev.ru\\
\end{center}

\begin{abstract}\noindent
In this paper a theory of reflective X-ray multilayer structures with a graded (slowly varying) period based on the coupled waves method and quasi-classical asymptotic expansions is reported. A number of exact solutions of the coupled wave equations is obtained and analyzed demonstrating suitability of this method for the description of the reflective properties of the graded multilayers. The developed theory is then used as a basis for the solution of the inverse problem, i.e. designing multilayer structures with a pre-specified reflectivity dependence on the wavelength or grazing angle. A number numerical experiments is conducted to demonstrate the capabilities of the proposed method in designing reflective multilayer coatings with an arbitrary shape of the reflectivity curve. The problem of maximization of the integral reflectivity is considered and a second order differential equation, which solutions correspond to multilayer structures with the maximal reflectivity, is derived. Finally, an upper limit on the integral reflectivity achievable with a graded multilayer is estimated.
\end{abstract}

\medskip


\section{Introduction} 
\pst
Multilayer reflective coatings are widely used in focusing and imaging X--ray optical elements and systems \cite{spiller1994soft,attwoodsoft}. Quite often they are simple periodical structures of two alternating materials containing from tens to thousands layers. Such periodic coatings are highly spectrally and angularly selective reflecting X-ray radiation only in a narrow range of wavelengths or grazing angles \cite{kozhevnikov1987basic}.

However, various applications of the reflective X-ray optics including those in the X-ray astronomy \cite{panini2018development,ragozin2009spectroscopic}, soft X-ray microscopy \cite{artioukov2004soft,artyukov2009soft} and EUV/BEUV lithography \cite{stulen1999extreme} would be much better served by the broadband aperiodic reflective multilayer coatings having extended reflectivity bands both in the wavelength and angular domains. This advantage is related to the ability of such optical elements to effectively utilize the radiation coming from broadband X-ray and EUV sources such free electron lasers, which could also significantly improve the spatial resolution and reduced the speckle artifacts. In many applications it is also important to optimize the multilayer coatings so that they have the maximal integral reflectivity in a broad spectral or angular passband rather than have the maximal peak reflectivity at a specific wavelength or grazing angle \cite{pirozhkov2015aperiodic,hofstetter2011lanthanum}.

It should be noted that for multilayer coatings the requirements of the high peak reflectivity and wide reflectivity band are in natural contradiction to each other. Finding a satisfactory compromise requires some sort of analytical theory, which is capable of solving the inverse problem for non-periodical multilayer coatings. Such a theory can be then used for the designing reflective coatings with an arbitrary dependence of the reflectivity on the wavelength or grazing angle as well as for the maximization of their integral reflectivity. A popular alternative to the analytical approach is the use of various numerical techniques including, for instance, the fastest descent method \cite{uspenskii2007optimal,pirozhkov2015aperiodic}, genetic algorithms and needle method \cite{tikhonravov2013design} for the optimization and inverse problem solution. On the other hand, over decades there have been considerable developments in the analytical description of non-periodic X-ray multilayer structures, which have met with varying success leaving a plenty of room for further improvements \cite{joensen1995design,protopopov1998x,kozhevnikov2015wideband}. 

In this paper we are building on our previous works \cite{vinogradov1998theory,vinogradov1999theory,vinogradov2000approach} and present a theory of X-ray multilayer coatings with the graded period, where \enquote{graded} means that the multilayer period significantly changes at a distance much larger than the period itself. There are two reasons to consider the graded multilayers: (i) they permit the accurate theoretical analysis, which can then be used as a starting point for the solution of the inverse problem, e.g. designing multilayers with the pre-specified reflectivity curve; (ii) manufacturing the multilayers with a gradually changing period is in practice a more realistic task as compared to fabrication of the multilayers having abrupt jumps of the period. Based on the developed theory we then solve the inverse problem for the multilayer coatings and propose a method to maximize their integral reflectivity.

\section{Coupled wave equations}
\pst
\label{sect_coupled}
Propagation and reflection of X-rays from a multilayer structure is assumed to obey the macroscopic electrodynamics. In this paper we consider the X-ray waves whose electrical field vector is perpendicular to the plane of incidence (s-polarization). However, all the results remain valid for the p-polarization if the grazing angle $\theta$ is small (grazing incidence) or close to $\pi/2$ (normal incidence). The important case when $\theta\approx\pi/4$ is beyond our consideration. We will also neglect the roughness of interfaces and other distortions of ideally sharp boundaries of layers in a multilayer.

Then the problem can be reduced to solution of the scalar Helmholtz equation in the following form
\begin{equation}
\frac{d^2E}{dz^2}+k^2\left[\varepsilon(z)-\cos^2\theta\right]E=0,
\label{fa1}
\end{equation}
where $E$ is the electric field, $z$ is the distance from the top of a multilayer, $\varepsilon(z)$ is the space distribution of the dielectric permittivity. 

For the analytical description of a graded multilayer with ideally sharp interfaces we use the following spatial dependence of the dielectric permittivity
\begin{equation}
\label{fa}
\varepsilon(z)=\tilde{\mu_{0}}+4\tilde{B}f(s),\quad s=2\int\limits_{0}^{z}q(z')dz',
\end{equation}
where
$$
\tilde{\mu_{0}}=\frac{\varepsilon_{1}+\varepsilon_{2}}{2},\quad
\tilde{B}=\frac{\varepsilon_{1}-\varepsilon_{2}}{8},
$$
$q(z)$ is the so called inverse period function, $\varepsilon_{1}, \varepsilon_{2}$ are dielectric constants of the two alternating multilayer materials and $f(s)$ is an even $2\pi$ periodical function, which can be written as
\begin{equation}
\label{fa5}
f(s)=1-2\sum\limits_{n=-\infty}^{\infty}\left[\theta(s-\pi\beta-2\pi n)-\theta(s-\pi(2-\beta)-2\pi n)\right],
\end{equation}
where $\theta(x)$ is a step-like theta function, $\beta=d_{1}/(d_{1}+d_{2})$ is a fraction of the period occupied by the more optically dense component 1 and $d_{1}$ and $d_{2}$ are the thicknesses of layers 1 and 2, respectively.

The dielectric function of a periodical structure can be obtained by assuming $q(z)=const$ in equation (\ref{fa}). The period of the multilayer structure is then $d=\pi/q$.

The Fourier series of $f(s)$ has the form
\begin{equation}
\label{fb}
\varepsilon(z)=\mu_{0}+4\sum_{n=1}^{\infty}B_{n}\cos(2n\int\limits_{0}^{z}q(z')dz'),
\end{equation}
where
$$
\mu_{0}=\eta\varepsilon_{1}+(1-\beta)\varepsilon_{2},\quad B_{n}=\frac{\varepsilon_{1}-\varepsilon_{2}}{2n\pi}\sin\pi n\beta.
$$
Substituting \eqref{fb} into the Helmholtz equation \eqref{fa1} we obtain
\begin{equation}
\label{fc}
E''+k^{2}\left[\mu+4B\cos(2\int\limits_{0}^{z}q(z')dz')\right]E=0,
\end{equation}
where $B=B_{1}$, $\mu=\mu_{0}-\cos^{2}\theta$. In \eqref{fc} following the papers \cite{spiller1994soft,kozhevnikov1987basic,vinogradov1977x} we kept only the first spatial harmonic of $\varepsilon(z)$ in \eqref{fb}, which is justified if the wavelength is close to the first Bragg resonance.

To solve equation (\ref{fc}) let us represent $E(z)$ as a sum of two waves with slowly varying amplitudes
\begin{equation}
\label{fd}
 E(z)=\sqrt{\frac{q(0)}{q(z)}}\left(b_{+}(z)e^{i\int\limits^{z}_{0}q(z')dz'}+b_{-}(z)e^{-i\int\limits^{z}_{0}q(z')dz'}\right).
\end{equation}
The coupled wave equations for the amplitudes $b_{\pm}(z)$ are obtained by substituting (\ref{fd}) into (\ref{fc}) and omitting the second derivatives of $b_{\pm}(z)$ as well as the terms with double phases (for more details see \cite{vinogradov1977x}), which lead to the following system
\begin{equation}
\label{fe}
\left\{
\begin{array}{rcl}
b_{+}'&=&i\frac{k^{2}}{2q(z)}\left(\mu-\frac{q^{2}(z)}{k^{2}}\right)b_{+}+i\frac{k^{2}B}{q(z)}b_{-},\\
b_{-}'&=&-i\frac{k^{2}B}{q(z)}b_{+}-i\frac{k^{2}}{2q(z)}\left(\mu-\frac{q^{2}(z)}{k^{2}}\right)b_{-}.
\end{array}
\right.
\end{equation}
The multilayer reflectivity $R$, transmittivity $T$ and the boundary conditions for $b_{\pm}(z)$ are
\begin{align}
R&=b_{-}(0),\quad b_{+}(0)=1,\label{fe1}\\
b_{-}(\zeta)&=0,\quad T=b_{+}(\zeta),\quad \zeta\to+\infty.\label{fe2}
\end{align}
The energy conservation law for the system \eqref{fe} is expressed by the relation
\begin{equation}
\label{fe3}
|b_{+}(z)|^{2}-|b_{-}(z)|^{2}=|T|^{2}=t,
\end{equation}
which follows directly from it and the boundary conditions \eqref{fe1}--\eqref{fe2} assuming that there is no absorption in the multilayer materials.

It can be shown that the amplitude $b=b_{-}(z)$ satisfies a second order differential equation, which follows from (\ref{fe})
\begin{equation}
\label{fh}
 b''+\frac{q'}{q}b'+\left[-iq'-\frac{k^{2}}{2}\mu+\frac{k^{4}}{4}\frac{(\mu^{2}-4B^{2})}{q^{2}}+\frac{q^{2}}{4}\right]b=0, 
\end{equation}
where $b=b_{-}(z)$ and the boundary conditions are
\begin{equation}
\label{fh1}
b(+\infty)=0,\quad b'(0)+i\frac{{k}^{2}}{2{q}}\left(\mu-\frac{{\rm q}^{2}}{{\rm k}^{2}}\right) b(0)=-i\frac{{k}^{2}{B}}{{q}}.
\end{equation}
Finally, the reflectivity $R$ can be expressed in terms of $b(z)$ as
\begin{equation}
\label{fj}
R=i\frac{k^{2}}{q(0)}B\frac{1}{\displaystyle i\frac{k^{2}}{q(0)}\left[\frac{q^{2}(0)}{k^{2}}-\mu\right]-\frac{b'(0)}{b(0)}}.
\end{equation}

\section{Exact models of graded multilayers}
\pst
\label{exact_form}
While for an arbitrary $q(z)$ the equation \eqref{fh} cannot be solved analytically, for a number of specific functions $q(z)$ the analytical solution is possible. In this section three such cases will be discussed: $q(z)=q_0(pz+\rho)$, $q(z)=q_0(pz+\rho)^{1/3}$ and $q(z)=q_0\tanh(pz+\rho)$, where $p$, $q_0$ and $\rho$ are some parameters.

\subsection{Linear case}
\pst
In the case when $q(z)=q_0(pz+\rho)$ one can always assume that $\rho=1$. Then introducing a new variable
\begin{equation}
\label{ch3ex1}
y=-\frac{iq_0}{2p}(pz+1)^2
\end{equation}
and representing the amplitude $b(z)$ as
\begin{equation}
\label{ch3ex2}
b(z)=Cy^\nu e^{-y/2}F(y),
\end{equation}
where $C$ is an arbitrary coefficient, it is possible to transform the equation \eqref{fh} into
\begin{equation}
\label{ch3ex3}
y\frac{d^2F}{dy^2}+[(2\nu+1)-y]\frac{dF}{dy}-\left[\nu+\frac{ik^2\mu}{4pq_0}\right]F=0,
\end{equation}
where
\begin{equation}
\nu=\frac{ik^2}{4pq_0}\sqrt{4\mu^2-B^2}.\label{ch3ex4b}
\end{equation}
Equation \eqref{ch3ex3} is the confluent hypergeometric equation \cite{andrews1999special}, which in general is known in the following form 
\begin{equation}
\label{ch3ex5}
y\frac{d^2F}{dy^2}+(c-y)\frac{dF}{dy}-a F=0,
\end{equation}
where $a$ and $c$ some complex coefficients. One of solutions of \eqref{ch3ex5} is the confluent hypergeometric function of the first type or Kumer function $M(a,c,y)={}_1F_1(a,c,y)$ \cite{andrews1999special}. However, it can be shown that the boundary condition \eqref{fh1} will be satisfied for a solution in the form \eqref{ch3ex2} only if the following solution of \eqref{ch3ex5} is used 
\begin{equation}
\label{ch3ex5a}
U_1(a,c,y)=-\frac{\pi}{\sin\pi c}\left[\frac{M(a,c,y)}{\Gamma(c)\Gamma(1-a)}-(-y)^{1-c}\frac{M(1+a-c,2-c,y)}{\Gamma(2-c)\Gamma(c-a)}\right].
\end{equation}
In the end the required solution of \eqref{fh} can be written as
\begin{equation}
\label{ch3ex6}
b(z)=Cy^\nu e^{-y/2}U_1(a,c,y),
\end{equation}
where
\begin{align}
a&=\frac{ik^2}{4pq_0}(\sqrt{\mu^2-4B^2}+\mu),\label{ch3ex7a}\\
c&=2\nu+1=\frac{ik^2}{2pq_0}\sqrt{\mu^2-4B^2}+1.\label{ch3ex7b}
\end{align} 
Now the reflectivity $R$ of the multilayer structure with a linearly growing inverse period $q$ can be obtained using the formula \eqref{fj}. Substituting the solution \eqref{ch3ex6} into it leads to the following expression
\begin{equation}
\label{ch3ex8}
R=i\frac{k^{2}}{2pq_0}B\frac{1}{\displaystyle a-y_0\frac{U_1(a+1,c+1,y_0)}{U_1(a,c,y_0)}},
\end{equation}
where $y_0=-iq_0/2p$. Equation \eqref{ch3ex8} can be simplified with use of the known properties of the confluent hypergeometric functions \cite{andrews1999special} and reduced to the following formula
\begin{equation}
\label{ch3ex9}
R=i\frac{k^{2}B}{2pq_0}\frac{U_1(a,c,y_0)}{U_1(a+1,c,y_0)},
\end{equation}
which can then be used for practical calculations of the reflectivity.

\subsection{Cubic root case}
\pst
In the case when $q(z)=q_0(pz+\rho)^{1/3}$, similar to the linear case it can be assumed without limiting the generality that $\rho=1$. Then the solution of \eqref{fh} can be similarly expressed through the confluent hypergeometric function. To do this let us introduce the following new variables
\begin{align}
y&=\sqrt{\frac{3q_0}{2p}}\left((pz+1)^{2/3}-\frac{k^2}{q_0^2}\mu\right),\label{ch3ex10a}\\
b(z)&=Ce^{iy^2/4}U(y).\label{ch3ex10b}
\end{align}
Substituting \eqref{ch3ex10a}--\eqref{ch3ex10b} into the equation \eqref{fh} reduces it to the so called the parabolic cylinder (Weber) equation 
\begin{equation}
\label{ch3ex11}
\frac{d^2U}{dy^2}+(y^2/4-a)U=0,
\end{equation}
where
\begin{equation}
\label{ch3ex10c}
a=\frac{i}{2}+\frac{3}{2}\frac{k^4}{pq_0^3}B^2.
\end{equation}
Solutions of the equation \eqref{ch3ex11} are Weber functions, which can be expressed through the confluent hypergeometric function \cite{andrews1999special}. Similar to the linear case, it can be shown that only the function $U_1$ defined in \eqref{ch3ex5a} has the correct asymptotic when $z\to\infty$. 

Thus the required solution of the equation \eqref{fh} and the respective reflectivity can be written as
\begin{align}
b(z)&=Ce^{iy^2/4}U_1(\alpha,1/2,-iy^2/2),\label{ch3ex12a}\\
R&=i\frac{k^{2}B}{q_0^2}\frac{1}{\displaystyle 1-\frac{k^2}{q_0^2}\mu}\frac{U_1(\alpha,1/2,-iy_0^2/2)}{U_1(\alpha+1,3/2,-iy_0^2/2)},\label{ch3ex12b}
\end{align}
where 
\begin{align}
\alpha&=\frac{ia}{2}+\frac{1}{4}=\frac{3i}{4}\frac{k^4}{pq_0^3}B^2,\label{ch3ex13a}\\
y_0&=\sqrt{\frac{3q_0}{2p}}\left(1-\frac{k^2}{q_0^2}\mu\right).\label{ch3ex13b}
\end{align}
Formula \eqref{ch3ex12b} is in many respects similar to the formula \eqref{ch3ex9} for the linear case. 

It is of interest to quantify the behavior of the obtained solution when $p\to0$. This can be done using the WKB (quasiclassical) asymptotic of the equation \eqref{ch3ex5} (see the next section). Their application in \eqref{ch3ex12b} leads to the following approximate expression for the reflectivity
\begin{equation}
\label{ch3ex13с}
R\approx\frac{R_0+iP}{1+iR_0 P},
\end{equation}
where $R_0$ is the reflectivity of the periodical multilayer structure with $q=q_0$ as defined in \cite{kozhevnikov1987basic}, while the coefficient $P$ is
\begin{equation}
\label{ch3ex13d}
P=R_0^{-i\frac{3q_0}{2p}}.
\end{equation}
Since it is known that inside the resonant reflectivity band $R_0=\exp(-i\phi)$, where the phase $\phi$ has a positive value, the coefficient $P$ will go to zero when $p\to0$, and $R\to R_0$. In other words, the coupled wave approximation provides for the transition from a graded multilayer to a periodical reflective structure. 

For the linear case discussed in the previous subsection an analogous transition is possible.

\subsection{Hyperbolic tangent case}
\pst
Finally let us consider the case when the inverse period function is $q(z)=q_0\tanh(pz+\rho)$. The main difference with the prior two models is the fact that the function $q$ assumes only the values from a limited interval $[q_0\tanh\rho,q_0]$. To solve the equation \eqref{fh} let us introduce the following variables
\begin{align}
y&=\tanh^2(pz+\rho),\label{ch3ex14a}\\
b(z)&=Cy^{-i\zeta}(1-y)^{\xi}U(y),\label{ch3ex14b}
\end{align}
where
\begin{align}
\zeta&=\frac{k^2}{4pq_0}\sqrt{\mu^2-4B^2},\label{ch3ex15a}\\
\xi&=\frac{k^2}{4pq_0}\sqrt{4B^2-\left(\mu-\frac{q_0^2}{k^2}\right)^2}.\label{ch3ex15b}
\end{align}
Substituting the formulas \eqref{ch3ex14a}--\eqref{ch3ex14b} into the equation \eqref{fh} the Gauss's hypergeometric equation is obtained \cite{andrews1999special}
\begin{equation}
\label{ch3ex16}
y(1-y)\frac{d^2U}{dy^2}+\left(c-(a+b+1)y\right)\frac{dU}{dy}-abU=0,
\end{equation}
where 
\begin{align}
a&=1-\frac{1}{4p}\left(iq_0+\frac{ik^2}{q_0}\sqrt{\mu^2-4B^2}-\frac{k^2}{q_0}\sqrt{4B^2-\left(\mu-\frac{q_0^2}{k^2}\right)^2}\right),\label{ch3ex17a}\\
b&=\frac{1}{4p}\left(iq_0-\frac{ik^2}{q_0}\sqrt{\mu^2-4B^2}+\frac{k^2}{q_0}\sqrt{4B^2-\left(\mu-\frac{q_0^2}{k^2}\right)^2}\right),\label{ch3ex17b}\\
c&=1-\frac{ik^2}{2pq_0}\sqrt{\mu^2-4B^2}.\label{ch3ex17c}
\end{align}
It can be shown that in order to obtain the correct asymptotic \eqref{fh1} when $z\to\infty$ the solution of \eqref{ch3ex16} should be chosen in the form $U(y)={}_2F_1(a,b,1+a+b-c,1-y)$, where ${}_2F_1$ is the Gauss's hypergeometric function. This results in the following expressions for the field amplitude and reflectivity
\begin{align}
b(z)&=Cy^{-i\zeta}(1-y)^{\xi}{}_2F_1(a,b,1+a+b-c,1-y),\label{ch3ex18a}\\
R&=i\frac{k^{2}B}{q_0}\frac{1}{\displaystyle i\frac{k^2}{2q_0}\left(\frac{q_0^2}{k^2}\tanh^2\rho-\mu\right)+Q},\label{ch3ex18b}
\end{align}
where
\begin{multline}
\label{ch3ex19}
Q=\left(i\frac{k^2}{q_0}\sqrt{\mu^2-4B^2}(1-\tanh^2\rho)+p(c'-1)\tanh^2\rho\right)+\\
2p\frac{ab}{c'}\tanh^2\rho(1-\tanh^2\rho)\frac{{}_2F_1(a+1,b+1,c'+1,1-\tanh^2\rho)}{{}_2F_1(a,b,c',1-\tanh^2\rho)},
\end{multline}
and
\begin{equation}
\label{ch3ex20}
c'=1+a+b-c=1+i\frac{k^2}{2pq_0}\sqrt{\left(\mu-\frac{q_0^2}{k^2}\right)^2-4B^2}.
\end{equation}
It is useful to consider the limit of \eqref{ch3ex18b} when $p\to\infty$. Taking into account \eqref{ch3ex17a}--\eqref{ch3ex17c} and the Taylor expansion of ${}_2F_1$ \cite{andrews1999special} it can be shown that in this case ${}_2F_1(a+1,b+1,2+a+b-c,1-\tanh^2\rho)\to1/\tanh^2\rho$ and ${}_2F_1(a,b,1+a+b-c,1-\tanh^2\rho)\to1$. Then the formula \eqref{ch3ex18b} for reflectivity turns into the know expression for the reflectivity of the periodic multilayer structure with the period $d=\pi/q_0$ (see \cite{kozhevnikov1987basic}).

The dependence of the inverse period in the form of the hyperbolic tangent from this subsection will be used in the numerical experiments of the section \ref{sect_WKB}, which are aimed at verifying the suitability of the coupled wave equations \eqref{fe} for the description of the graded multilayers. 

\section{WKB asymptotic for the coupled wave equations}
\label{sect_WKB}

\begin{figure}[t!]
\includegraphics[scale=0.3]{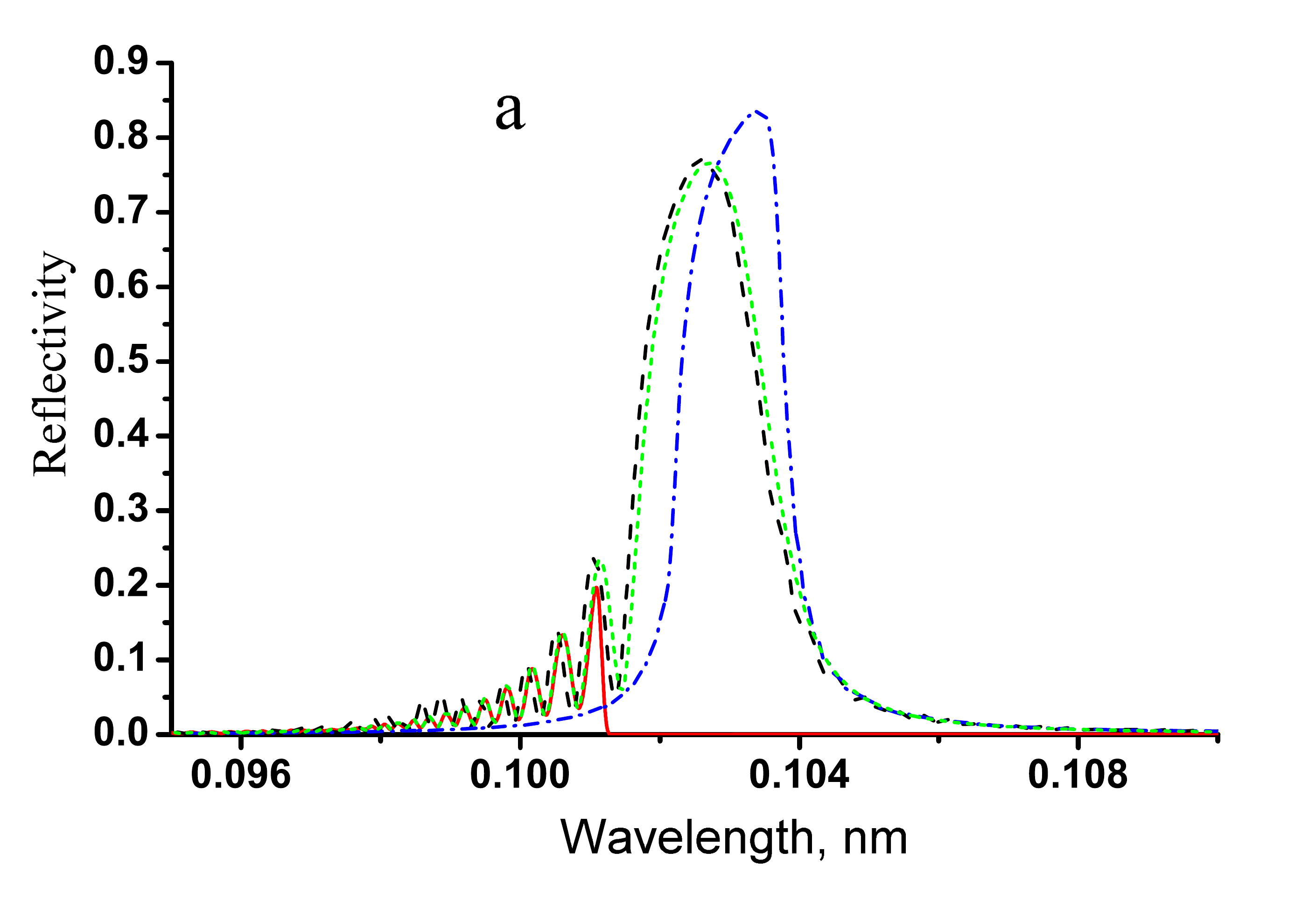}
\includegraphics[scale=0.3]{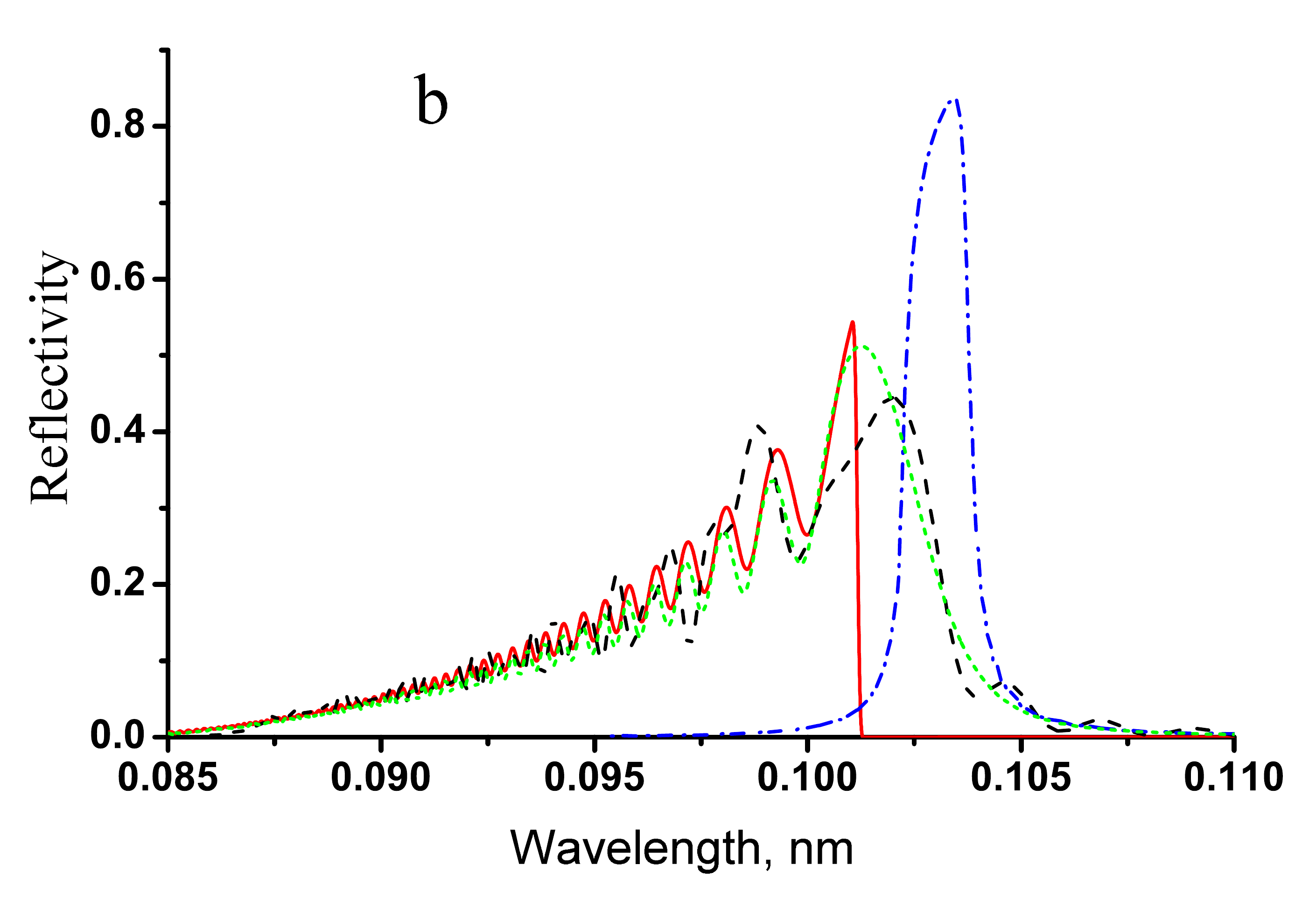}

\includegraphics[scale=0.3]{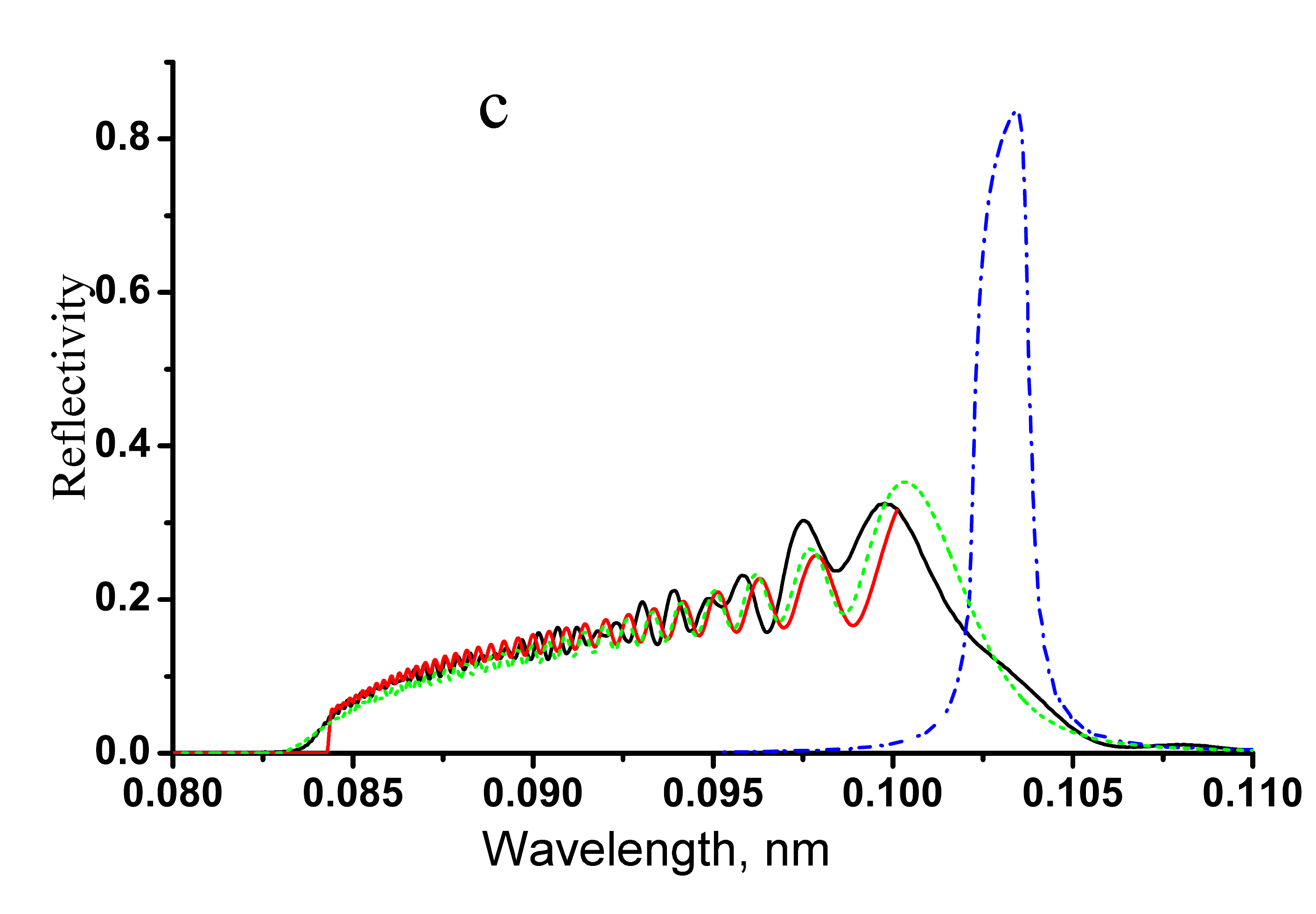}
\includegraphics[scale=0.3]{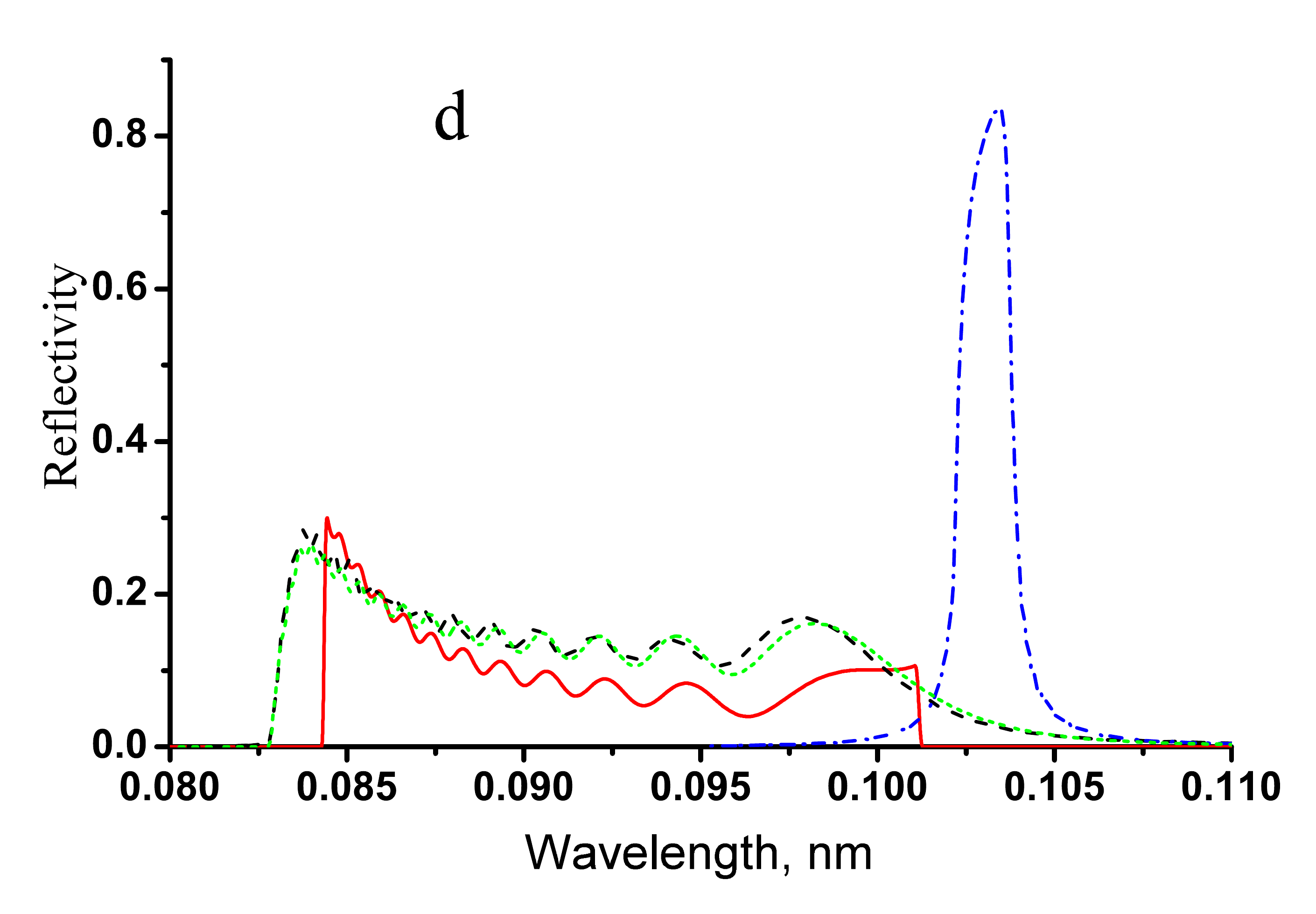}
\caption{Comparison of the reflectivities calculated using: the approximate asymptotic formula (\ref{fs}), the analytical formula \eqref{ch3ex18b} and the matrix method. The multilayer structure is $Cr/C$ with $q(z)=q_{0}\tanh(pz+\rho)$, where $q_{0}=1.3$ $\mbox{nm}^{-1}$, $\rho=1.1$, $\beta=0.5$ and $\theta=1^{\circ}$. The black dash line corresponds to the matrix method, green short-dash line -- to the analytical formula, red solid line -- to the approximate asymptotic formula and the blue dash-dot line -- to the limiting periodic structure when $p\to0$. The panels correspond to the following values of $p$: a -- $p=10^{-4}$ $\mbox{nm}^{-1}$, b -- $p=5\cdot 10^{-4}$ $\mbox{nm}^{-1}$, c -- $p=1\cdot10^{-3}$ $\mbox{nm}^{-1}$, d -- $p=3\cdot10^{-3}$ $\mbox{nm}^{-1}$.}
\label{f1}
\end{figure}

\begin{figure}[t!]
\includegraphics[scale=0.3]{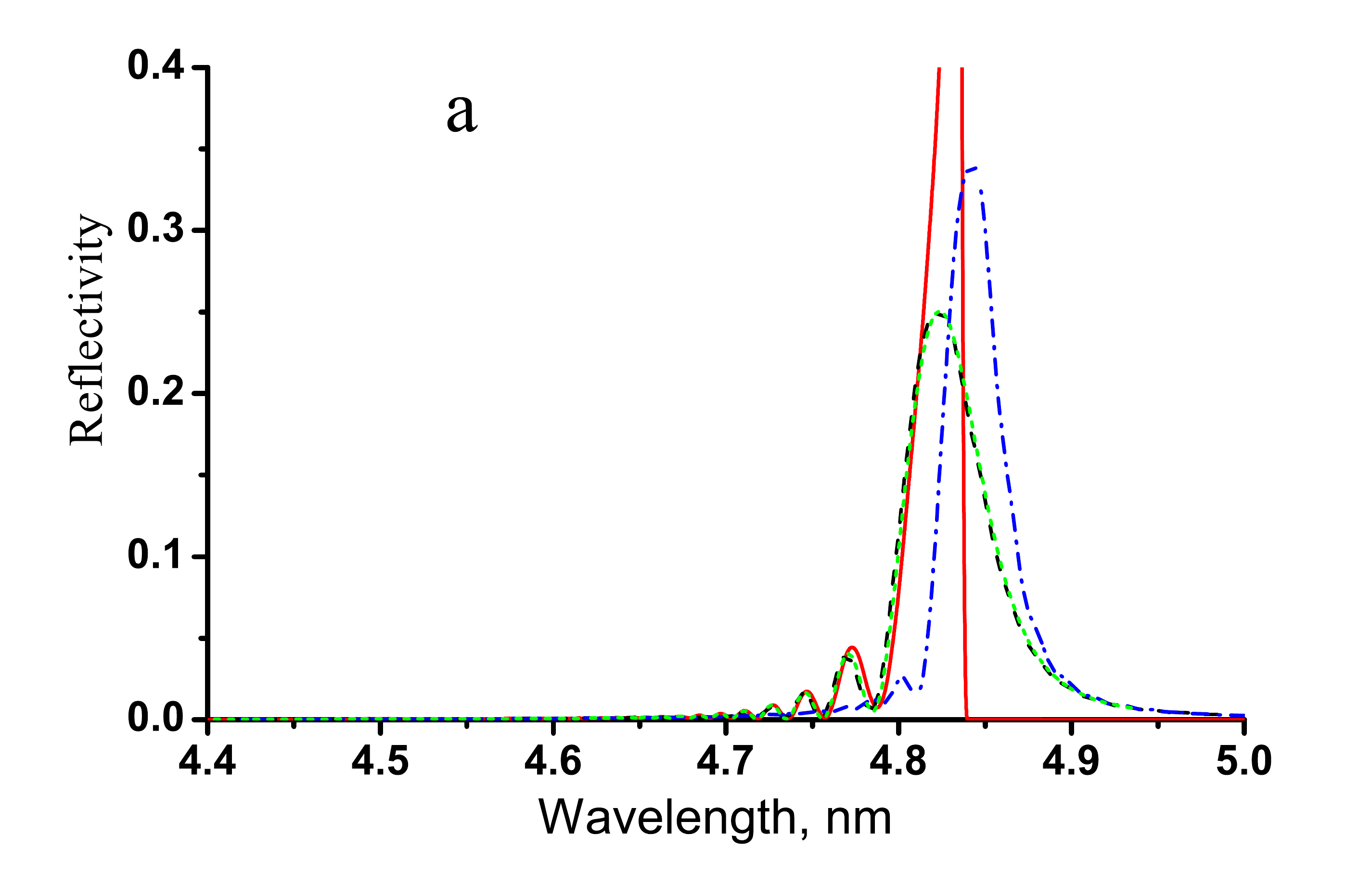}
\includegraphics[scale=0.3]{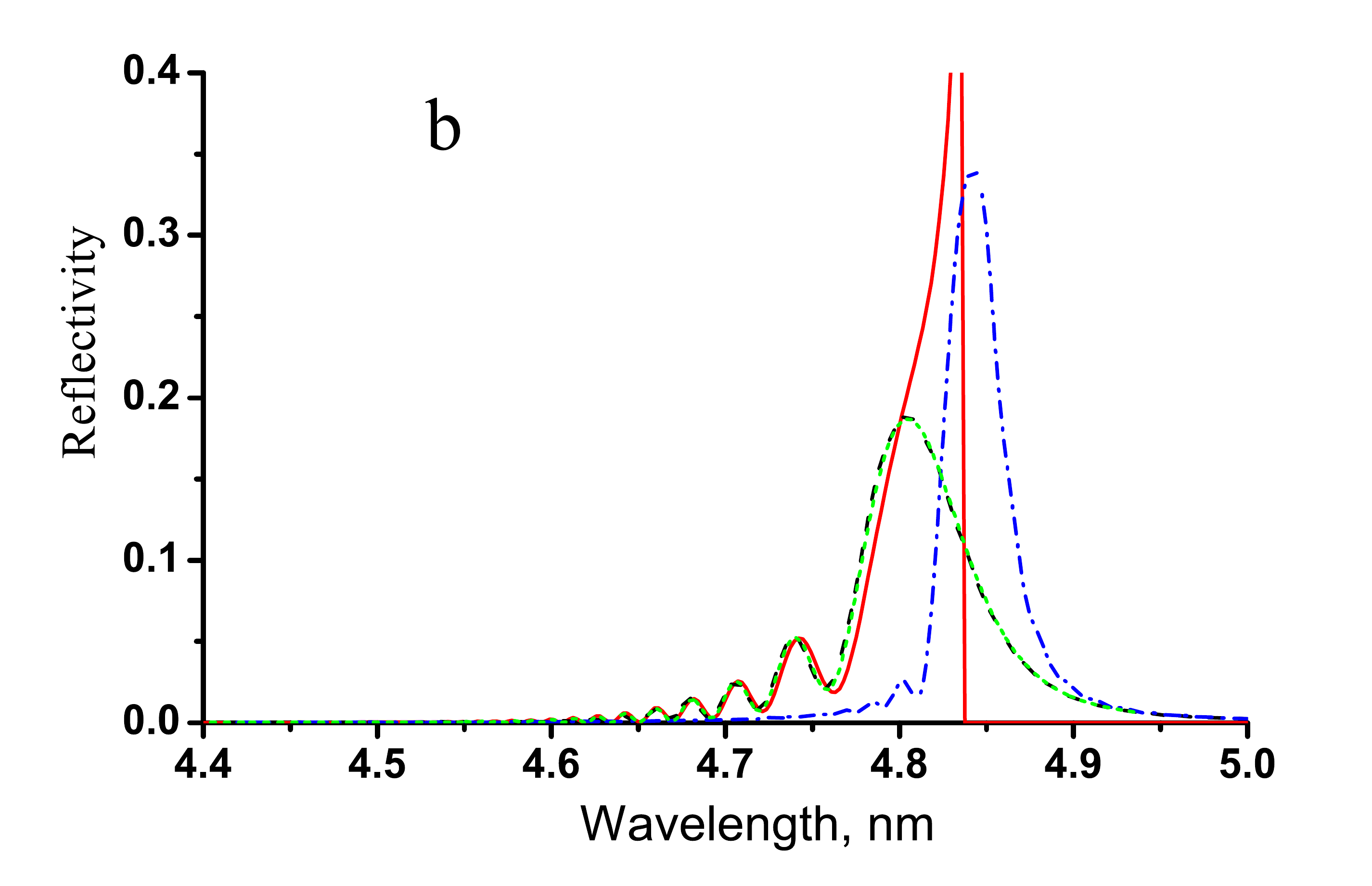}

\includegraphics[scale=0.3]{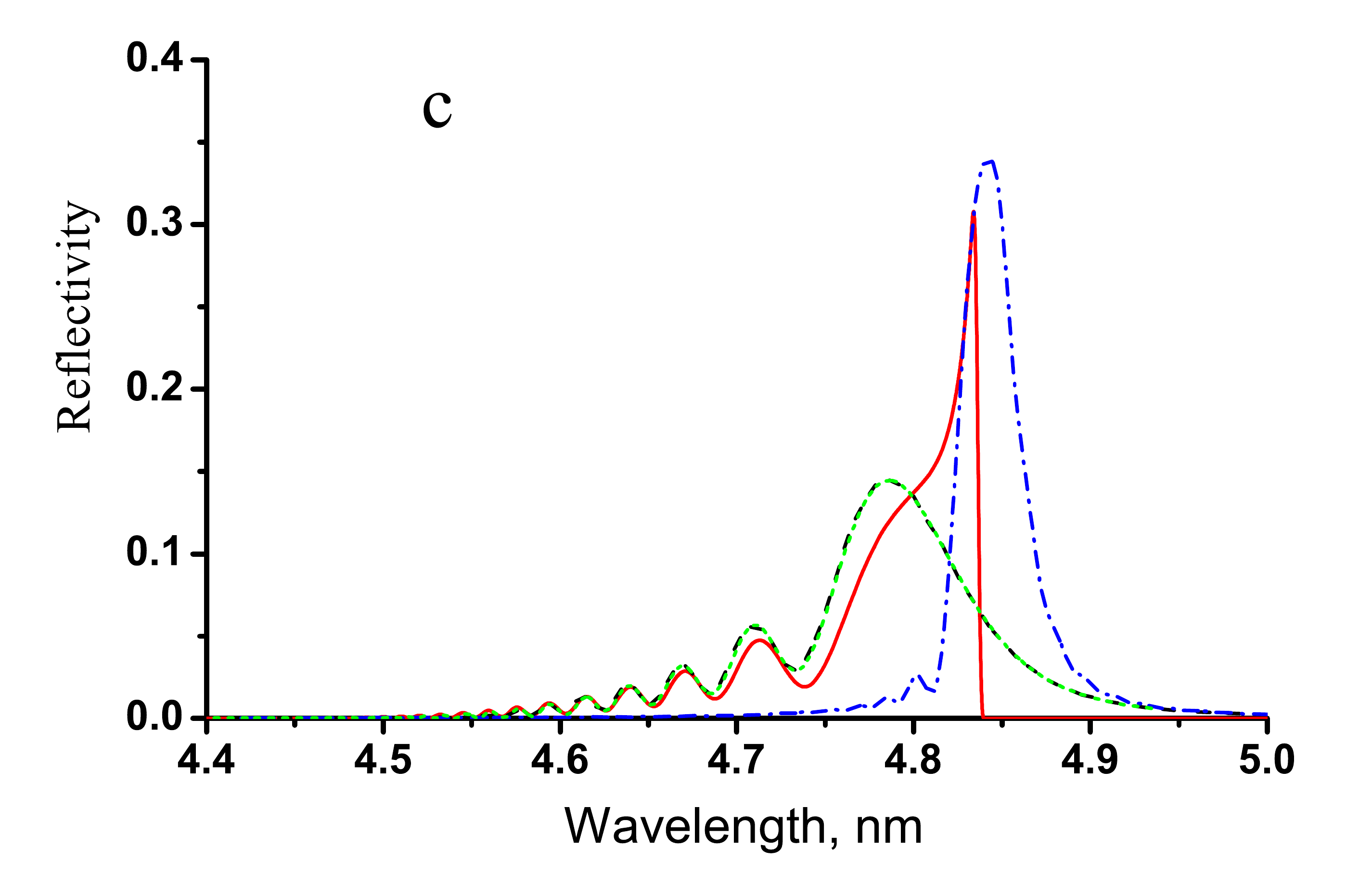}
\includegraphics[scale=0.3]{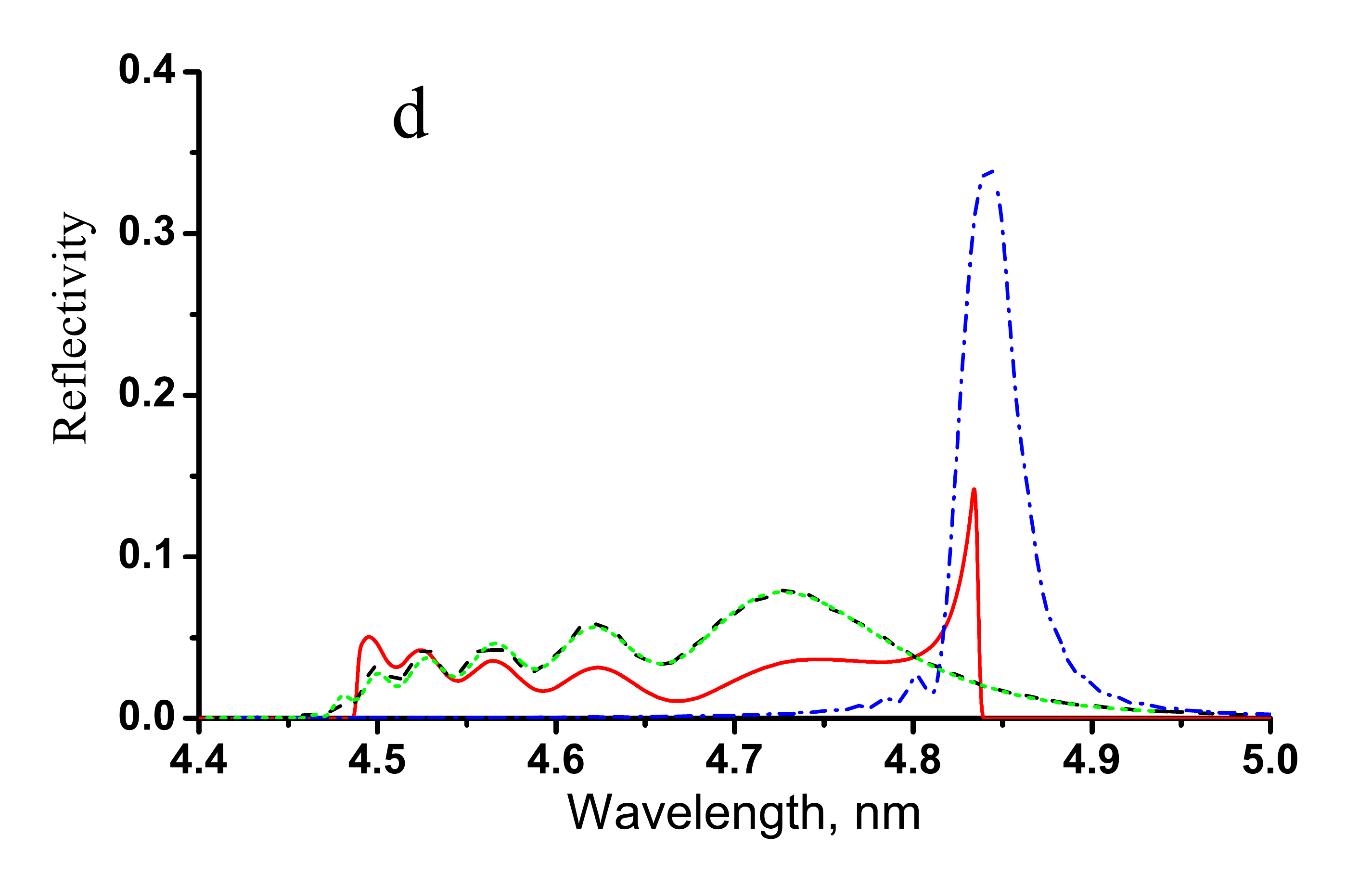}
\caption{Comparison of the reflectivities calculated using: the approximate asymptotic formula (\ref{fs}), the analytical formula \eqref{ch3ex18b} and the matrix method. The multilayer structure is $Cr/C$ with $q(z)=q_{0}\tanh(pz+\rho)$, where $\mbox{q}_{0}=1.4$ $\mbox{nm}^{-1}$, $\rho=1.58$, $\beta=0.4$ and $\theta=90^{\circ}$. The black dash line corresponds to the matrix method, green short-dash line -- to the analytical formula, red solid line -- to the approximate asymptotic formula and the blue dash-dot line -- to the limiting periodic structure when $p\to0$. The panels correspond to the following values of $p$: b -- $p=3\cdot10^{-4}$ $\mbox{nm}^{-1}$, b -- $p=6\cdot10^{-4}$ $\mbox{nm}^{-1}$, c -- $p=10^{-3}$ $\mbox{nm}^{-1}$, d -- $p=3\cdot10^{-3}$ $\mbox{nm}^{-1}$.}
\label{f2}
\end{figure}

Equation (\ref{fh}) is equivalent to the following equation
\begin{equation}
\label{fk}
\ddot{{y}}+\left[\frac{f_{0}(x)}{{p}^{2}}+\frac{f_{1}(x)}{{p}}+f_{2}(x)\right]{y}=0,
\end{equation}
which is the standard form for application of the WKB (Wentzel, Kramers, Brilloin) asymptotic method \cite{olver1997asymptotics}. In the equation \eqref{fk}
$$
\begin{array}{rcl}
f_{0}&=&\frac{{q}^{2}}{4}+\frac{{k}^{4}}{4}\frac{(\mu^{2}-4{B}^{2})}{{q}^{2}}-\frac{{k}^{2}}{2}\mu,\\
f_{1}&=&-i\dot{{q}},\\
f_{2}&=&\frac{1}{4}\left(\frac{\dot{{q}}}{{q}}\right)^{2}-\frac{1}{2}\frac{\ddot{{q}}}{{q}},
\end{array}
$$
where $z=px$, $y=\sqrt{q}b$ and $p$ is a formal parameter of the asymptotic expansion. The general solution of (\ref{fk}) can be expressed in the form of series by the parameter $1/p$ as
\begin{equation}
\label{fl}
y(z)=\exp{\left(\frac{1}{p}S_{0}(x)+S_{1}+pS_{2}(x)+\ldots\right)},
\end{equation}
where formulas for $S_{n}$ can be obtained by substituting \eqref{fl} into (\ref{fk}). By keeping only the first two terms in (\ref{fl}) we can obtain an approximate solution of the coupled wave equations in the following form
\begin{equation}
\label{fm}
b(z)=\frac{1}{\sqrt{q\gamma}}\left\{C_{1}\left[i(q^{2}+2\zeta)+2q\gamma\right]^{\frac{1}{2}}e^{\int\limits^{z}\gamma(z')dz'}+C_{2}\left[i(q^{2}+2\zeta)+2q\gamma\right]^{-\frac{1}{2}}e^{-\int\limits^{z}\gamma(z')dz'}\right\},
\end{equation}
where
$$
\zeta=-\frac{{k}^{2}}{2}\mu,\quad\gamma=\sqrt{-f_{0}},\quad\gamma'=\sqrt{f_{0}}.
$$
The solution (\ref{fm}) is only valid at large distances from the so-called turning points of equation \eqref{fk}, which coincide with the zeros $z_{i}$ of the function $f_{0}(z)$ at
\begin{equation}
\label{fn}
q=k\sqrt{\mu\pm2B}.
\end{equation}
For the sake of simplicity let us at first neglect the absorption and restrict the consideration to the real values of $z$. It can be shown that, if $q(z)$ changes monotonically, the equation \eqref{fn} has at most two solutions $z_{1}$ and $z_{2}$ lying on the positive real semi-axis. Moreover, from the point of view of the width of the reflectivity band the most important case is when both zeros are located inside the multilayer structure. This corresponds to the wavelength $\lambda$ being within the resonant part of the reflectivity band
\begin{equation}
\frac{2\pi\sqrt{\mu}}{q_1}<\lambda<\frac{2\pi\sqrt{\mu}}{q_2},
\label{fo}
\end{equation}
where it is assumed that $q_1>q_2$. The full WKB solution of (\ref{fk}) can be constructed by extending the general solution (\ref{fm}) from one area in the complex plane of $z$ to all other areas by using its expansions in the vicinity of the turning points $z_{1}$, $z_{2}$. Taking into account the boundary conditions \eqref{fh1} we obtain that when $z>z_{1}$
\begin{equation}
\label{fp}
b(z)=\frac{C}{\sqrt{{q}\gamma'}}\frac{e^{-i\int\limits_{z_{1}}^{z}\gamma'(z')dz'}}{\left[{A}+\sqrt{{A}^{2}-1}\right]^{1/2}},
\end{equation}
where
$$
A(z)=\frac{1}{2{B}}\left(\mu-\frac{{q}^{2}(z)}{k^{2}}\right).
$$
The expression for the amplitude $b(z)$ in the area where $z_{2}<z<z_{1}$ is
\begin{equation}
\label{fq}
 b({z})=\frac{C}{\sqrt{{q}\gamma}}\left\{\frac{M}{\left[{A}-i\sqrt{1-{A}^{2}}\right]^{1/2}}e^{\psi_{1}}+i\left(\frac{1}{M}-M\right)\left[{A}-i\sqrt{1-{A}^{2}}\right]^{1/2}e^{-\psi_{1}}\right\},
\end{equation}
where
$$
\psi_{1}=\int\limits_{z_{2}}^{z}\gamma(z')dz',\quad 
M=e^{-\int\limits_{z_{2}}^{z_{1}}\gamma(z')dz'}.
$$
Finally, the amplitude $b(z)$ in the area where $z<z_{2}$ is
\begin{equation}
\label{fr}
b({z})=\frac{C}{\sqrt{{q}\gamma'}}\left\{\frac{\left[{A}-\sqrt{{A}^{2}-1}\right]^{1/2}}{M}e^{-i\psi}+i\left(\frac{1}{M}-M\right)\frac{1}{\left[{A}-\sqrt{{A}^{2}-1}\right]^{1/2}}e^{i\psi}\right\},
\end{equation}
where
$$
\psi=-\int\limits_{z_{2}}^{z}\gamma'(z')dz'.
$$
By substituting the solutions (\ref{fp})--(\ref{fr}) into the general formula (\ref{fj}) we obtain the following expression for reflectivity inside the resonant reflectivity band (\ref{fo})
\begin{equation}
\label{fs}
R=\frac{\displaystyle 1+i\omega\frac{e^{2i\psi}}{{A}-\sqrt{{A}^{2}-1}}}{\displaystyle {A}+\sqrt{{A}^{2}-1}+i\omega e^{2i\psi}},
\end{equation}
where
\begin{eqnarray}
\psi&=&\int\limits_{0}^{z_{2}}\gamma'(z')dz',\nonumber\\
\omega&=&1-e^{-2\alpha},\label{fs1}\\
\alpha&=&\int\limits_{z_{2}}^{z_{1}}\gamma(z')dz',\nonumber\\
A&=&\frac{1}{2{B}}\left(\mu-{q}^{2}(0)/{k}^{2}\right).
\nonumber
\end{eqnarray}
Formula (\ref{fs}) gives the reflectivity of a graded multilayer in terms of the material's optical constants and multilayer geometry. It can be further simplified if one is interested only in the module of the reflectivity. In the resonant reflectivity band (\ref{fo}) band and when $|A|\gg1$ it can be shown that
\begin{align}
R\approx&\frac{1}{2{A}}+i\omega e^{2i\psi},\label{ft1}\\
r=&|R|^{2}\approx-\operatorname{Im}(\frac{1}{{A}}\omega e^{2i\psi})+|\omega|^{2}e^{-4\psi''},\label{ft2}
\end{align}
where
$$
\psi''=\operatorname{Im}(\psi).
$$
In \eqref{ft2} the first term describes the oscillations of the reflectivity and the second term corresponds to the average reflectivity. Taking into account that the reflectivity band of any graded multilayer is considerably wider than the reflectivity band of a comparable periodical structure and is determined primarily by the gradient of the function $q(z)$ rather than by the absorption, one can use the following estimates for the parameters $\alpha$ and $\psi''$
\begin{eqnarray}
\alpha&=&\pm\pi k^2\frac{{B}^{2}}{\operatorname{Re}(\mu)}\left|\left(\frac{dq}{dz}(\tilde{z})\right)^{-1}\right|,\label{fu}\\
\psi''&=&\beta\frac{k}{2\sqrt{\operatorname{Re}(\mu)}}\tilde{z},\label{fv}
\end{eqnarray}
where $\tilde z=q^{-1}(z)$ and $\mu=\sin^2\theta-\delta+i\beta$. This is equivalent of approximating the equation (\ref{fn}) for the zeros of $f_0(z)$ as $q(z)=k(\sqrt{\mu}\pm B/\sqrt{\mu})$. Let us note that in (\ref{fu}) and (\ref{fv}) it is also assumed that the absorption is small, i.e. that $\beta\ll\delta\ll 1$.

The formula \eqref{ft2} can be further simplified by averaging over the oscillations and by substituting $B$ for $|B|$. The latter can be done because the absorption is small. Then the parameter $\omega$ becomes a real number and for the averaged reflectivity we will have
\begin{equation}
r\approx\omega^{2}e^{-4\psi''}.
\label{fv1}
\end{equation}

Finally, to access the accuracy of the approximate formulas obtained in this section we conducted a number of numerical experiments. We chose the model function $q(z)=q_0\tanh(pz+\rho)$, where $q_0$, $p$ and $\rho$ are some parameters, and for which the equation \eqref{fh} has the exact solution as it was demonstrated above. Then we compared the reflectivity calculated using the formula \eqref{fs} with that obtained by the direct solution of the Helmholtz equation (\ref{fa1}) in a multilayer coating using the matrix method for the dielectric susceptibility dependence defined in \eqref{fa5}.

To apply the matrix method one has to express the bilayer thicknesses $d_n$ in terms of function $q(z)$. This can be done by using the definition of function $q(z)$ to obtain the positions of the bilayer interfaces $z_n$
\begin{equation}
2\int\limits_{0}^{z_{n}}q(z)dz=2\pi n,
\label{fz3}
\end{equation}
which are extremums of $\cos(2\int\limits_{0}^{z_{n}}q(z)dz)$ in (\ref{fb}). After that the bilayer thicknesses can be calculated as $d_n=z_n-z_{n-1}$, where $n=1,2,\ldots$.

All calculations with the analytical formulas as well as by the direct matrix method \cite{kozhevnikov1987basic} were performed using the specially written FORTRAN 95 program codes. The X-ray optical constants were taken from the Henke tables \cite{henke1993x}.

Two such comparisons for $Cr/C$ multilayer structures in the hard and soft (\enquote{carbon window}) X-ray spectral ranges are shown in Fig.\ref{f1}--\ref{f2}, which also depict the reflectivities calculated using the analytical formula \eqref{ch3ex18b}. As it is seen from Fig.\ref{f1}--\ref{f2} the approximate formula (\ref{fs}) is in a reasonably good agreement with the direct solution of the Helmholtz equation (\ref{fa1}) sometimes reproducing even small features of rather complicated profiles of $r(\lambda)$, although it diverges and is clearly inapplicable near the boundaries of the resonant reflectivity band. 

On the other hand, the formula \eqref{ch3ex18b} gives the results, which are even in a better agreement with the direct solution. The accuracy of the analytical formula increases when the gradient of the inverse period decreases ($p\to0$) or when the reflectivity band is narrow (as in Fig.\ref{f2}). This is exactly what is expected from the general theory formulated in the sections \ref{sect_coupled} and \ref{sect_WKB}. 

\section{Inverse problem}
\pst
\label{sect_inv}
\begin{figure}[!t]
\includegraphics[scale=0.3]{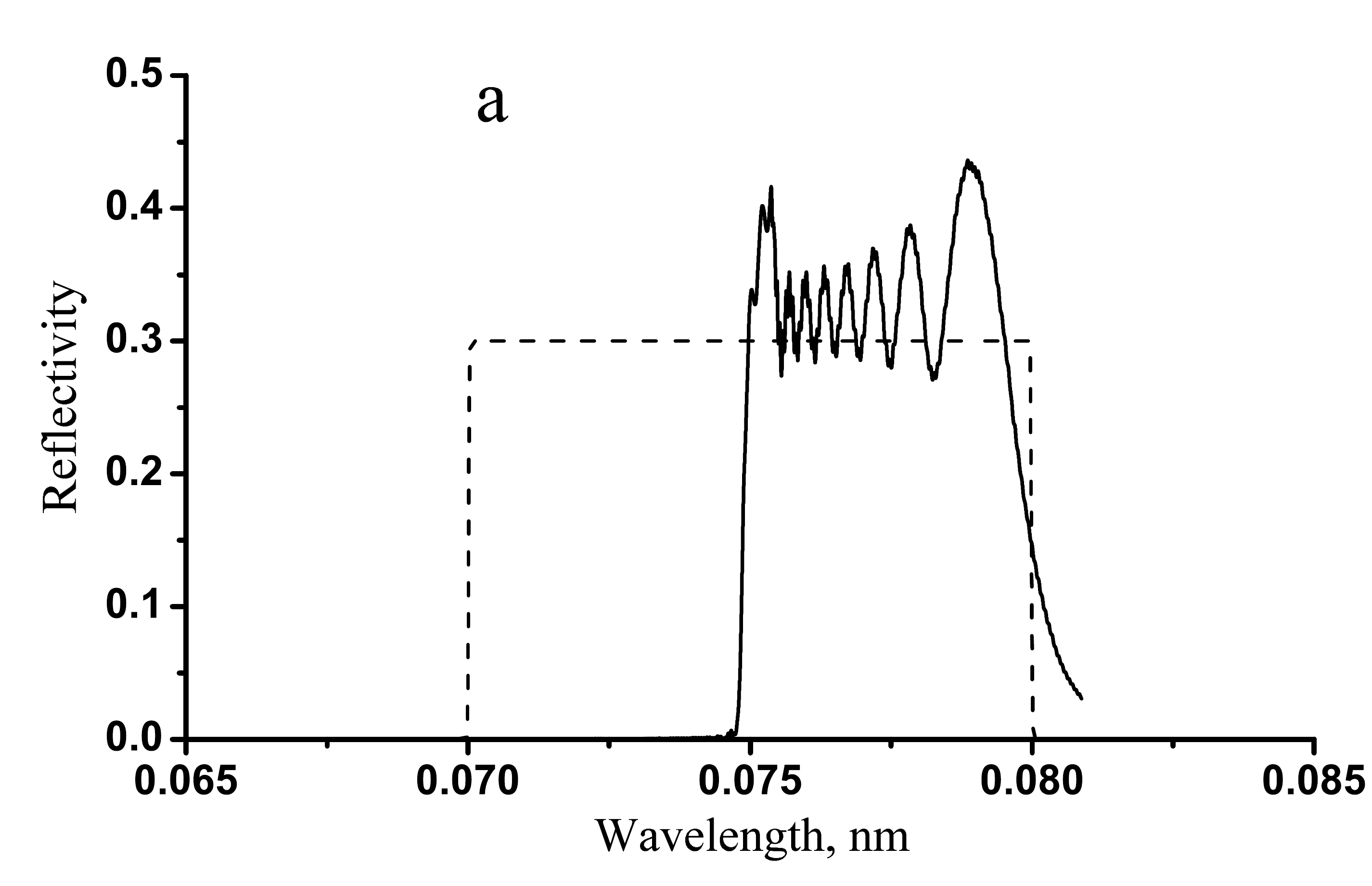}
\includegraphics[scale=0.3]{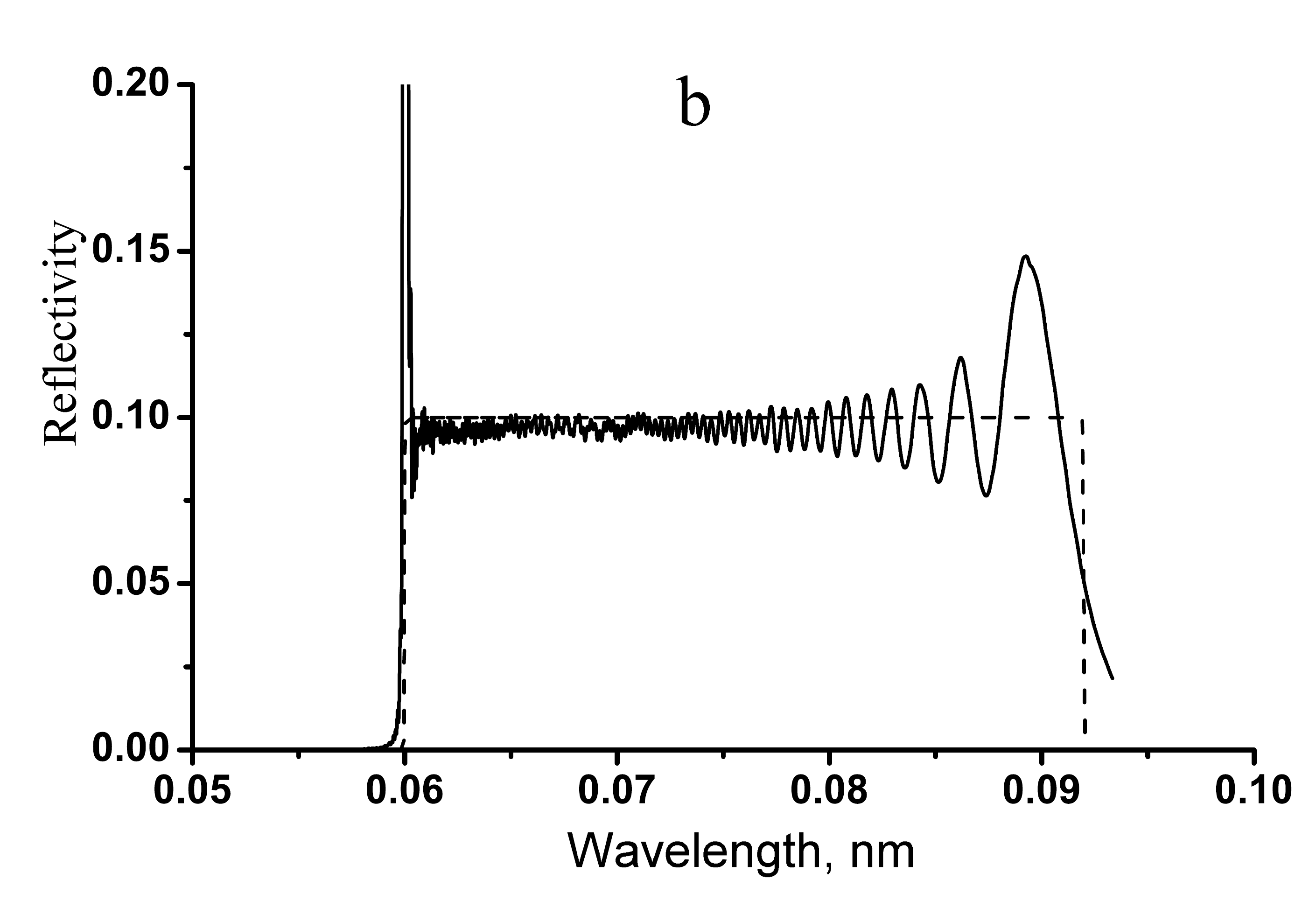}

\includegraphics[scale=0.3]{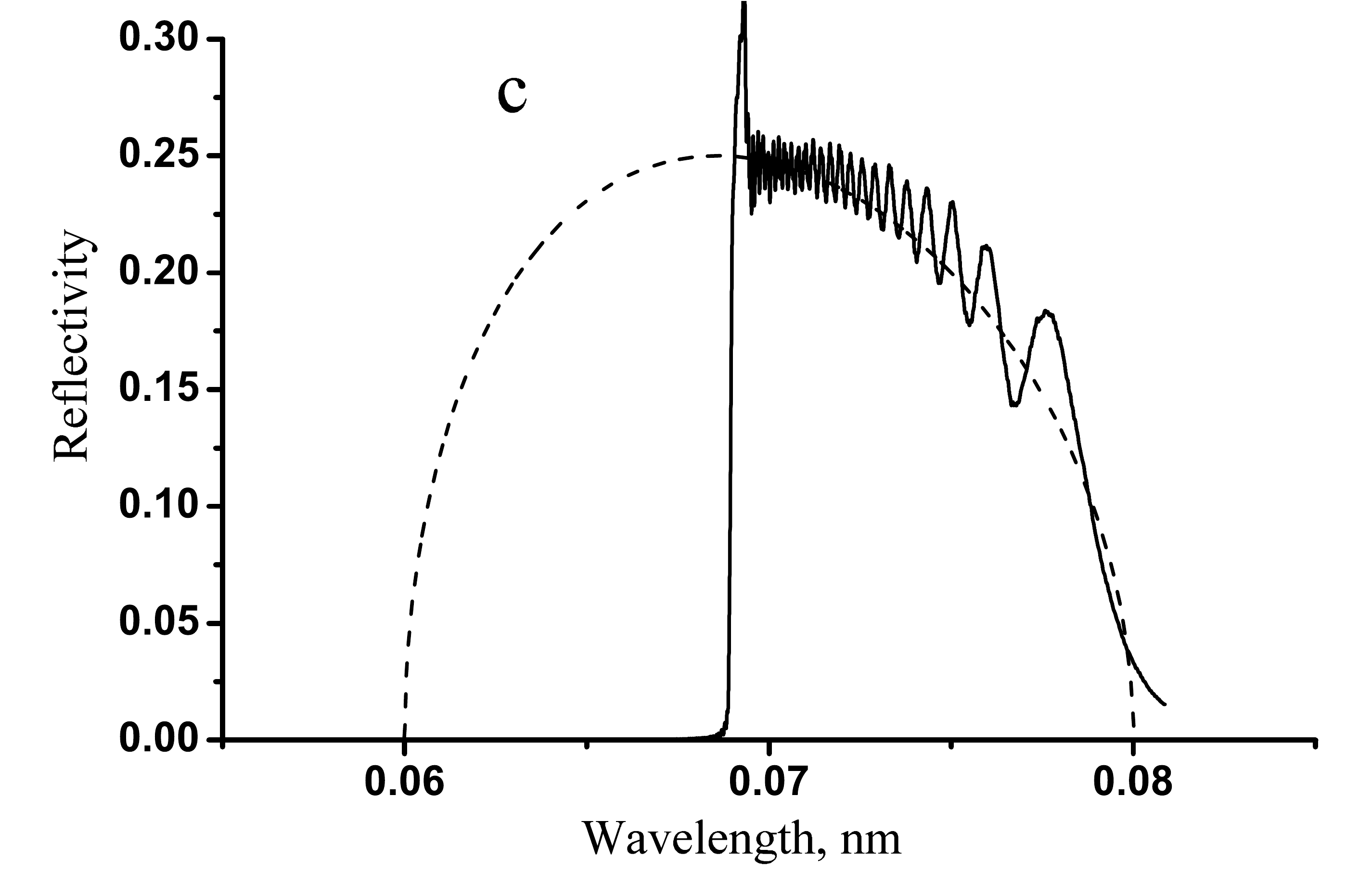}
\includegraphics[scale=0.3]{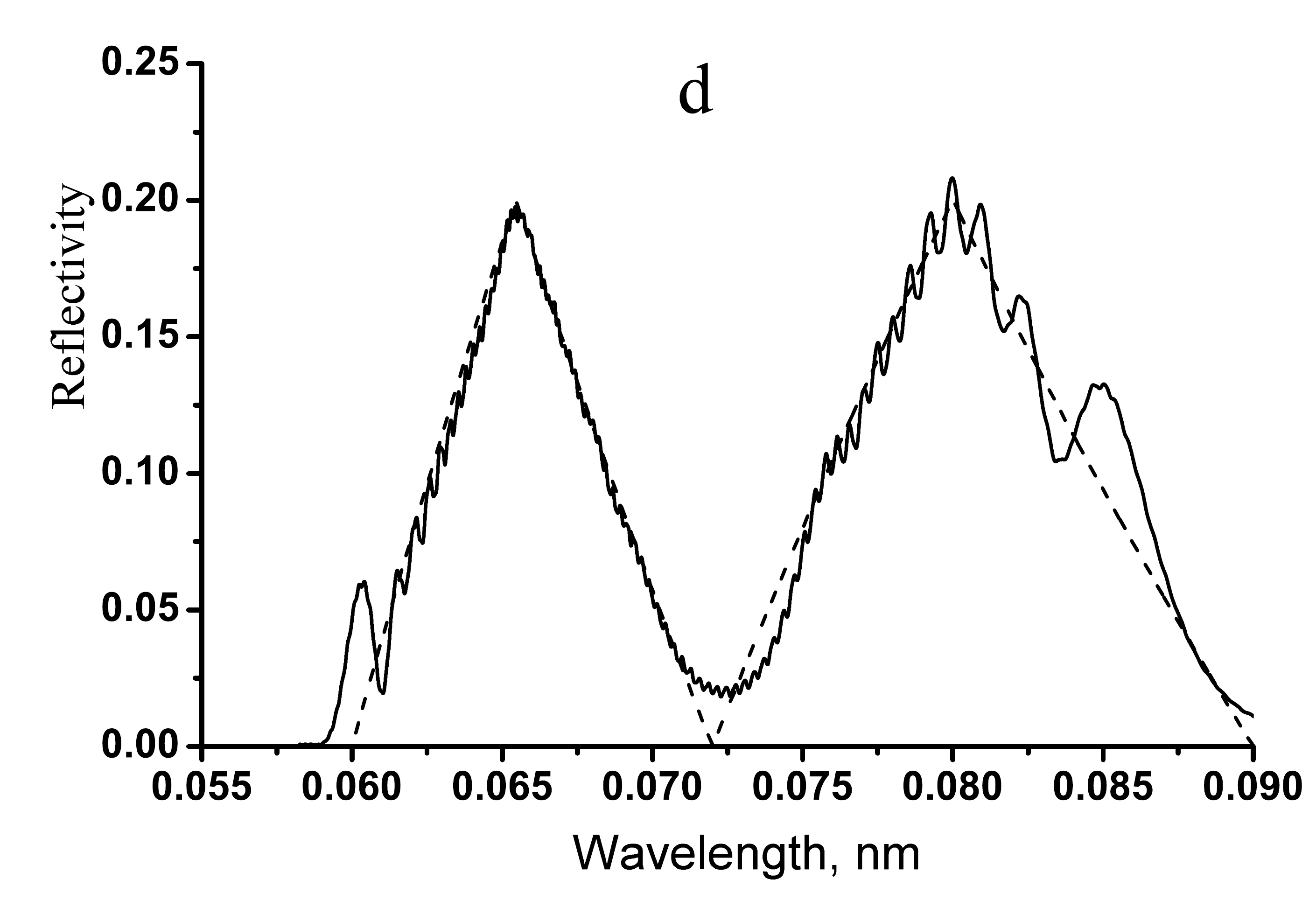}
\caption{The examples of the solution of the inverse problem in a hard X-ray range with $\lambda\approx0.1$ nm for the $Cr/C$ multilayer structure with $\beta=0.38$ and for the grazing angle $\theta=1^{\circ}$. The dash line is a pre-specified reflectivity and solid line is the reflectivity of the multilayer structure constructed using the method developed in this paper. The $q(z)$ functions for these four structures are depicted in Fig.\ref{f5}.}
\label{f3}
\end{figure}

\begin{figure}[!t]
\includegraphics[scale=0.3]{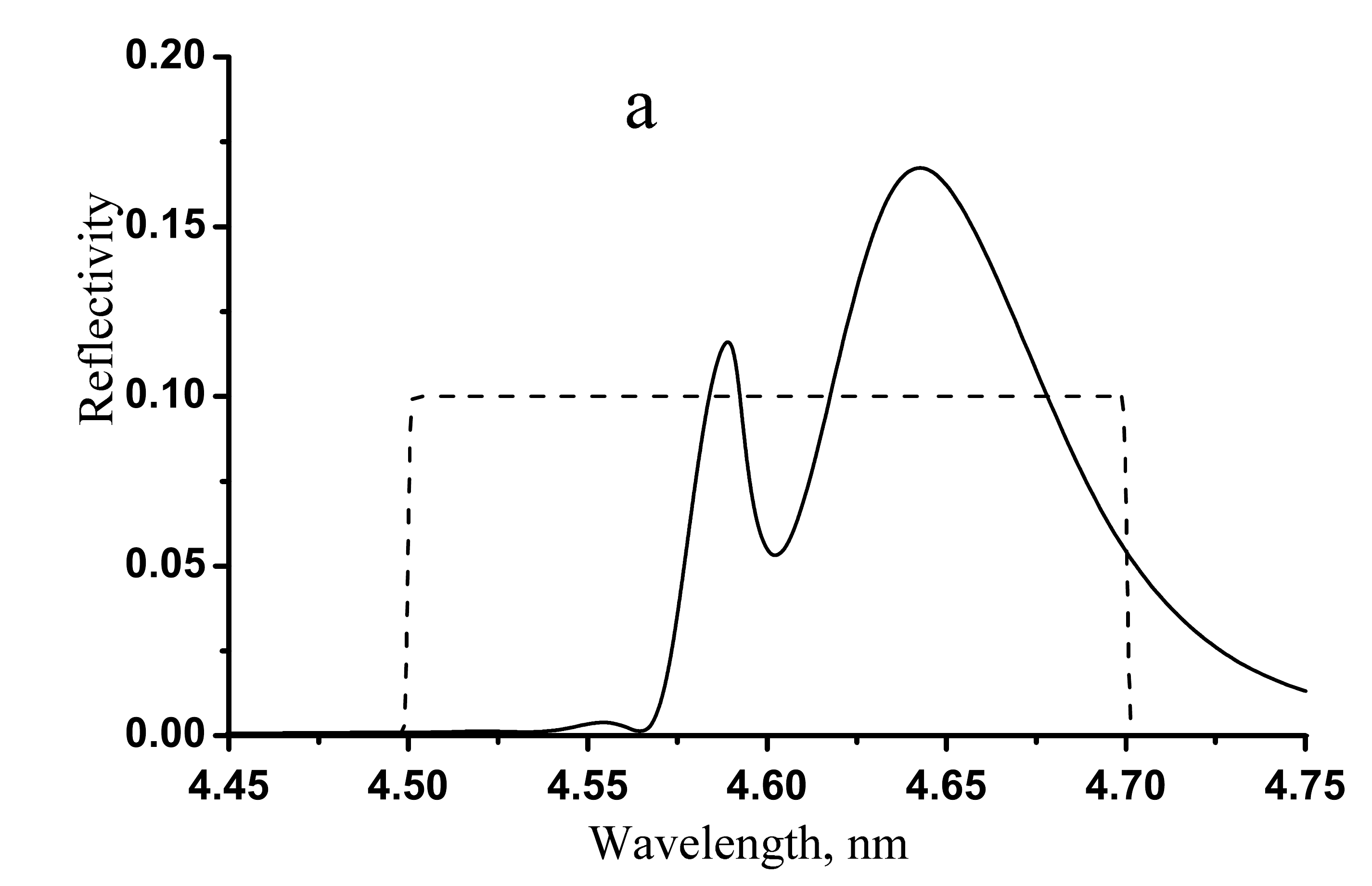}
\includegraphics[scale=0.3]{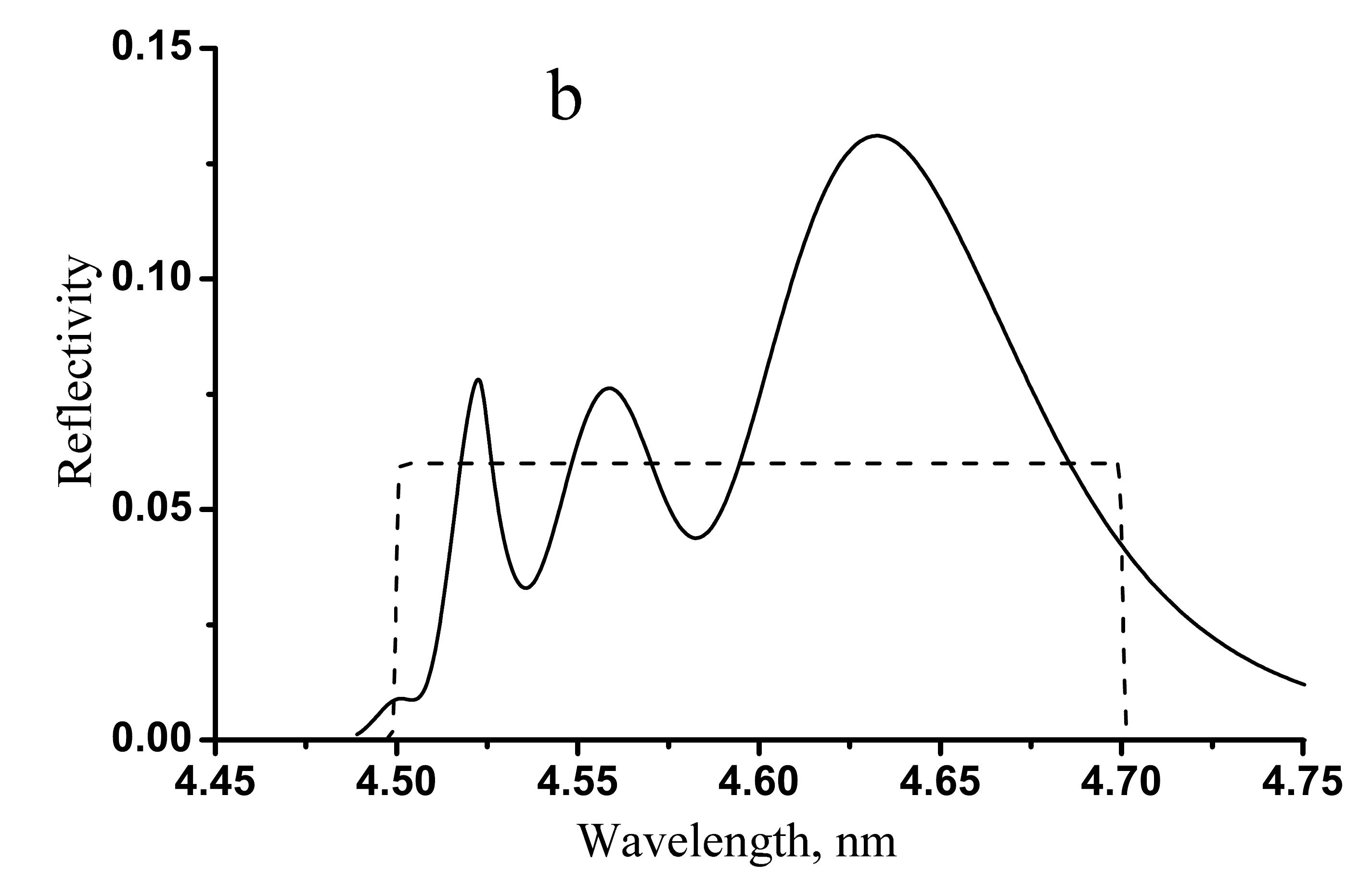}

\includegraphics[scale=0.3]{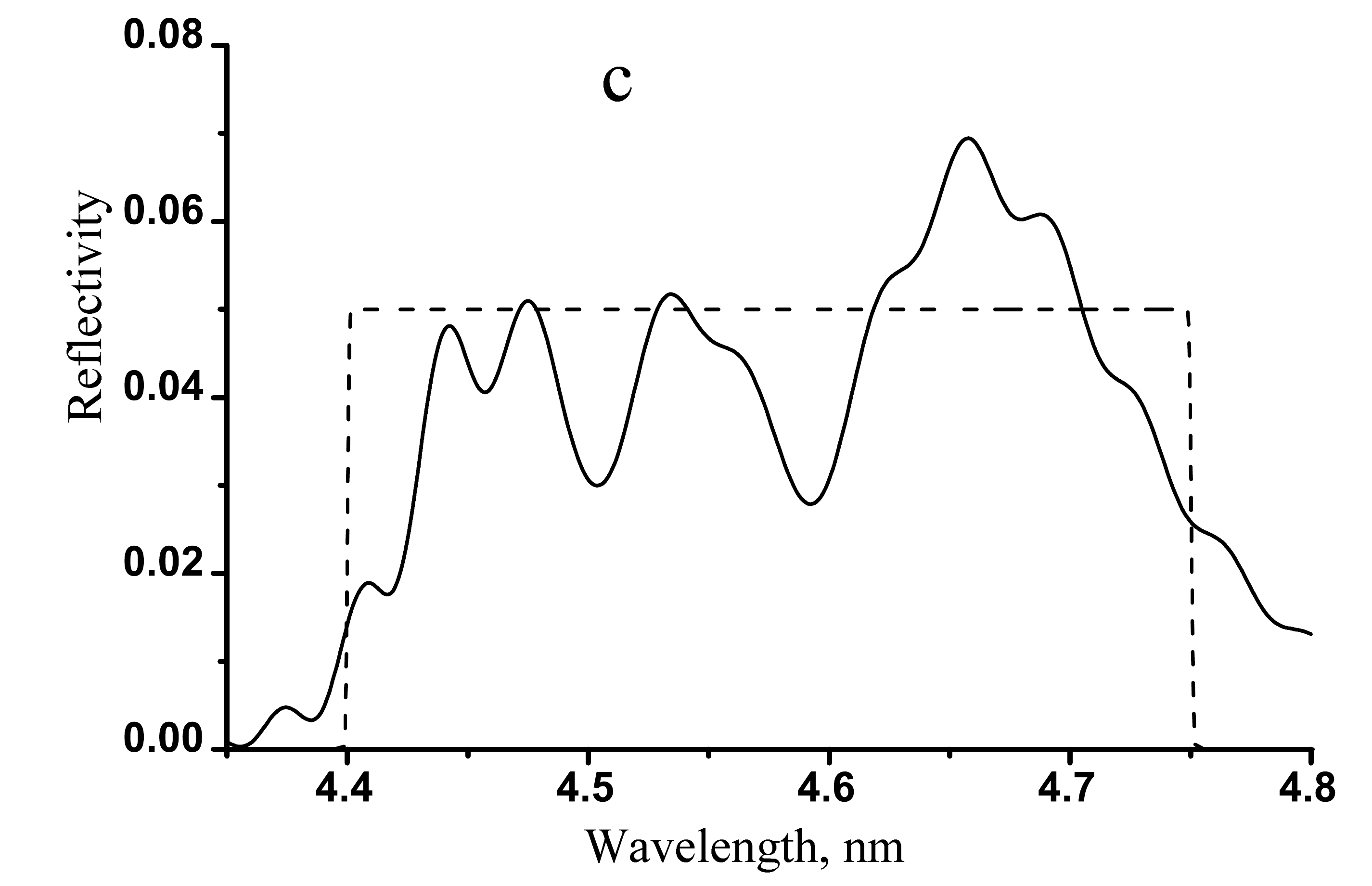}
\includegraphics[scale=0.3]{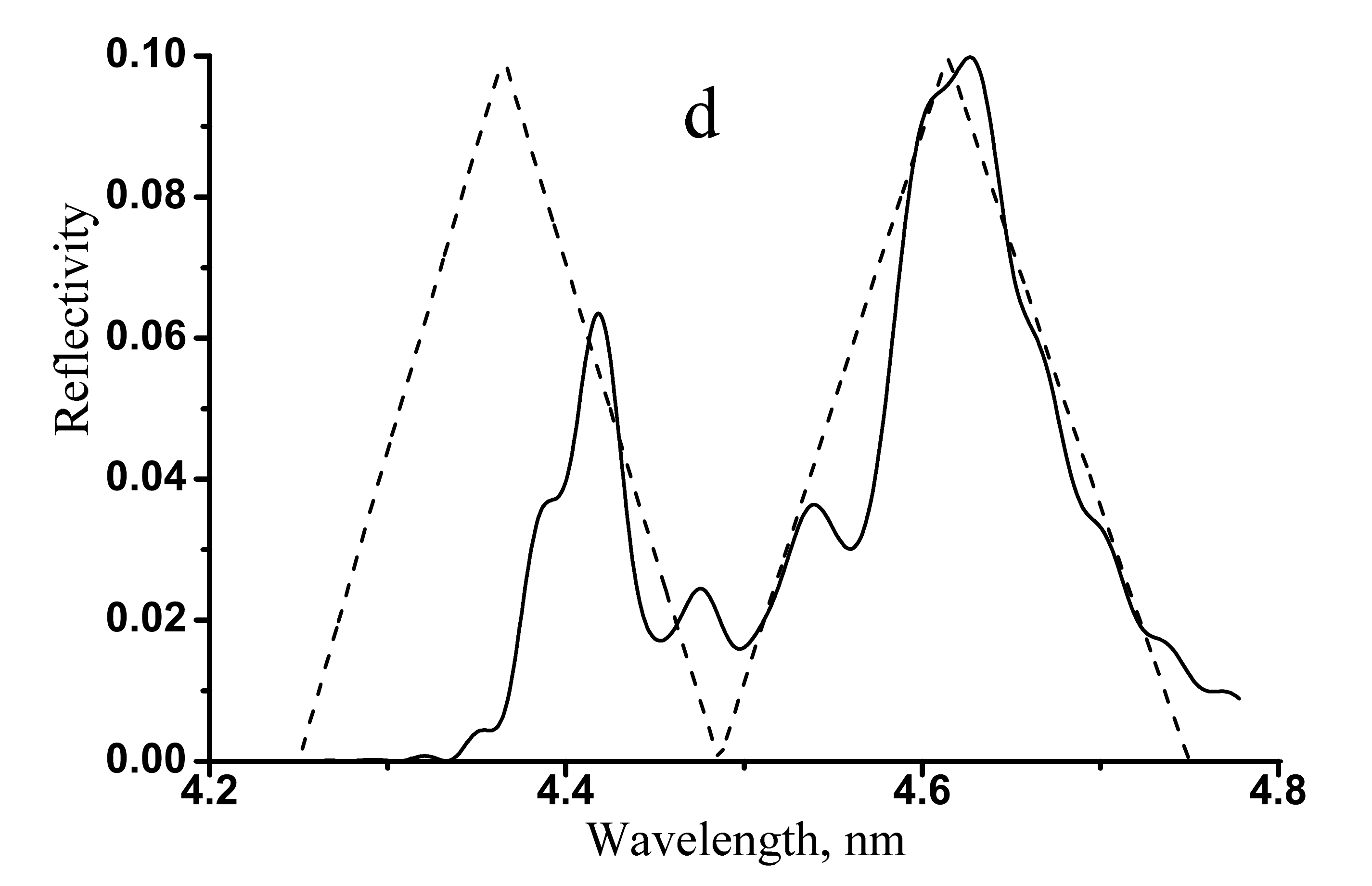}
\caption{The examples of the solution of the inverse problem in the \enquote{carbon window} range with $\lambda\approx4.4--5.0$ nm for the $Cr/C$ multilayer structure with $\beta=0.38$ and for the grazing angle $\theta=90^{\circ}$. The dash line is a pre-specified reflectivity and solid line is the reflectivity of the multilayer structure constructed using the method developed in this paper.}
\label{f4}
\end{figure}

It is very useful to compare results of the perturbation theory, which is also known as the kinematic approximation (see \cite{vinogradov1999theory} for details), and results given by the WKB asymptotics. Let us first observe that the formula (\ref{fv1}) in case $|B|\leqslant1$ behaves asymptotically as $r\approx4\alpha^{2}e^{-4\psi''}\sim B^{4}$. On the other hand, when $|B|\leqslant1$ the kinematic approximations is also applicable, from which it follows that the reflectivity should be proportional to $B^{2}$, but not to $B^{4}$, i.e. to the square of modulation of the dielectric permittivity in the multilayer structure (see \cite{vinogradov1999theory}). To rectify this problem we will use the following formula for the parameter $\omega$  instead of the expression \eqref{fs1}
\begin{equation}
\omega\approx\sqrt{\tanh(\alpha)}.
\label{fz4}
\end{equation}
In the limit when $\alpha\to\infty$ formula \eqref{fz4} coincides with \eqref{fs1} simultaneously making the reflectivity \eqref{fv1} behave correctly in the kinematic approximation, i.e. when $\alpha\to0$. 

The numerical experiments have also demonstrated that use of the formula (\ref{fv1}) may lead to the values of the reflectivity that are somewhat elevated over those obtained numerically by the matrix method. Since this can cause errors in the inverse problem solution the following modified formula for the reflectivity is used here
\begin{equation}
r\approx\frac{r_{old}}{1+r_{old}},
\label{fz5}
\end{equation}
where
$$
r_{old}=\tanh(\alpha)e^{-4\psi''},
$$
which alleviates this problem. 

Now the formula (\ref{fz5}) together with the Bragg condition $q(\tilde{z})=\sqrt{\mu}k$ can be used to derive a differential equation for the function $q(z)$, which solves the inverse problem of constructing a multilayer coating with a given dependence of the reflectivity on the wavenumber $k$ or grazing angle $\theta$. 

For the wavelength dependence it has the following form
\begin{equation}
\frac{dz}{dq}=\mp\frac{(\operatorname{Re}(\mu))^{2}}{2\pi |B|^{2}}\frac{1}{q^{2}}\ln\left(\frac{1-\frac{r}{1-r}e^{4\psi''}}{1+\frac{r}{1-r}e^{4\psi''}}\right) 
\label{fz2}
\end{equation}
with the initial condition $z(q_{0})=0$ and where
$$
r=r\left(\frac{q}{\sqrt{\operatorname{Re}(\mu)}}\right)
$$
and $r(x)$ is a pre-specified function of $k$. 

For the grazing angle dependence the differential equation is 
\begin{equation}
\frac{dz}{dq}=\mp\frac{q^2}{2\pi k^{4}}\frac{1}{B^2} \ln\left(\frac{1-\tilde r(q)(1+e^{4\psi''})}{1-\tilde r(q)(1-e^{4\psi''})}\right),
\label{fz6}
\end{equation}
with the same initial condition $z(q_{0})=0$ and where
$$
\tilde r(q)=r\left(\frac{q^2}{k^2}+1-\operatorname{Re}(\mu_0)\right)
$$
and $r(x)$ is a given function of $\sin\theta$. The plus sign in \eqref{fz2} and \eqref{fz6} should be chosen to obtain the increasing function $q(z)$, and the minus sign -- the decreasing function $q(z)$. In other words, every inverse problem has at least two solutions.

To demonstrate the ability of the developed method to solve real world inverse problems a number of numerical experiments in different wavelength ranges and with diverse shapes of the reflectivity curve has been carried out. Their results are shown in Fig.\ref{f3}--\ref{f4}. The differential equation \eqref{fz2} for the growing $q(z)$ was solved numerically by an explicit Euler method using a specially written FORTRAN 95 program code. The obtained functions $q(z)$ were converted to the layer thicknesses using the equation \eqref{fz3}, which then were used to calculate the multilayer reflectivities (also shown in Fig.\ref{f3}--\ref{f4}) by the direct matrix method \cite{kozhevnikov1987basic} as it was done in the numerical experiments of the previous section. The X-ray optical constants were again taken from the Henke tables \cite{henke1993x}.

It is clear from Fig.\ref{f3}--\ref{f4} that the method proposed here is capable of constructing blueprints of the graded multilayers, whose reflectivities approximate the target reflectivities reasonably well except for the observed quasiperiodic oscillations. The agreement is appreciably better in the hard X-ray region than in the soft \enquote{carbon window} spectral range. The calculations also showed (see Table \ref{t1}) that the graded multilayers might have the integrated (over wavelength) reflectivity at least several times larger than that of a periodical structure. Thus the developed theory can be used as a basis for the solution of the inverse problem and maximization of the integral reflectivity.

\begin{wrapfigure}{r}{0.5\textwidth}
\includegraphics[scale=0.25]{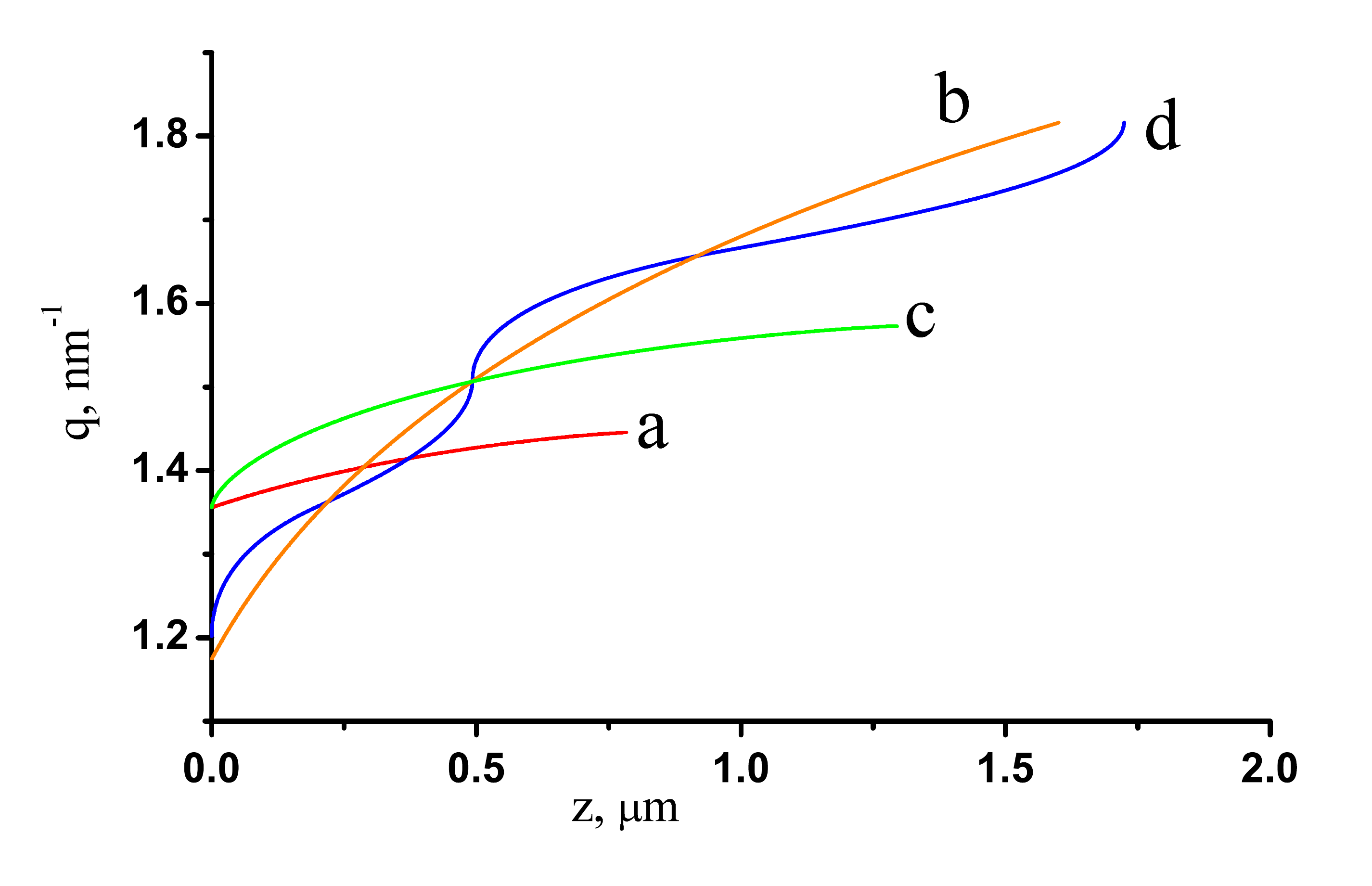}
\caption{The inverse period functions $q(z)$ for the multilayer structures, whose reflectivities are shown in Fig.\ref{f3}. The letters correspond to the panels in \ref{f3}.}
\label{f5}
\end{wrapfigure}

\section{Maximization of the integral}
\pst
The reflectivity in (\ref{fv1}) depends on the function $q^{-1}(q)$, which is the inverse of the function $q(z)$, and on its first derivative. Therefore we can identify the reflectivity $r$ with the Lagrange function known from the classical mechanics \cite{landau2013mechanics}, the integral reflectivity -- with the action, $q^{-1}(q)$ -- with a generalized coordinate and the wavelength or grazing angle with the time. The maximization problem for the integral reflectivity in an interval of wavelengths (or grazing angles) is then reduced to the Lagrange equation
\begin{equation}
\frac{\partial r}{\partial z}=\frac{d}{dq}\frac{\partial r}{\partial z'_{q}}
\label{ch3eq2h}
\end{equation}
with the boundary conditions
\begin{equation}
z(q_1)=0,\quad z(q_2)=L,\label{ch3eq2ha}
\end{equation}
where $q_1$ and $q_2$ are the boundary values of the function $q(z)$ and $L$ is the multilayer's thickness.

From the equation (\ref{ch3eq2h}) the following second order differential equation is obtained
\begin{multline}
\label{ch3eq2i}
\frac{d^2z}{dq^2}=\frac{1}{2a^3q^3}\frac{4c\frac{\tanh\xi}{1-\tanh^2\xi}\pm\frac{da}{dq}q\pm2a}{\tanh\xi+e^{-4\psi''}}(1+r_0)-\frac{1}{aq^2}\left(\frac{da}{dq}q^2\frac{dz}{dq}+2aq\frac{dz}{dq}\right)\mp\\
\frac{2}{aq^2}\frac{1-r_0}{\tanh\xi+e^{-4\psi''}}\left(\frac{dc}{dq}qz+cz+cq\frac{dz}{dq}\right),
\end{multline}
where the plus or minus sign corresponds to an increasing or decreasing function $q(z)$, respectively.

In the equation (\ref{ch3eq2i})
\begin{eqnarray}
r_0&=&\tanh\xi e^{-4\psi''},\quad \psi''=cqz,\nonumber\\
\xi&=&\pm aq^{2}\frac{dz}{dq},\quad \mu'=\operatorname{Re}(\mu),\nonumber\\
a&=&\pi\frac{|B|^2}{(\mu')^2},\quad c=\frac{\gamma}{2}\frac{1}{(\mu')^2}.\nonumber
\end{eqnarray}
The derivatives of $a$ and $c$ assume non-zero values only if the material dispersion is present. For the maximization of the integral over an angular interval an equation similar to \eqref{ch3eq2i} can be obtained. 

Equation \eqref{ch3eq2i} is quite complicated and in the general case can be solved only numerically. It is, however, possible to estimate the upper limit of the achievable integral reflectivity in rather a simple way. 

Let us assume that $R(\lambda)$ (or, respectively, $R(\theta)$) has a table-like profile in the wavelength interval $[\lambda_{min},\lambda_{max}]$ (or in the grazing angle interval $[\theta_{min},\theta_{max}]$)
\begin{equation}
r(\lambda)=r_c(\theta(\lambda-\lambda_{min})-\theta(\lambda-\lambda_{max})),
\label{ch3eq2g3}
\end{equation} 
where $r_c$ is a positive number less than 1. For \eqref{ch3eq2g3} the inverse problem can be solved as described in the section \ref{sect_inv}. The maximum or minimum value of $q$ will be determined by the location of zero in the argument of logarithm in \eqref{fz2} or \eqref{fz6}, respectively. After the maximal inverse period range $\Delta q_{max}=|q_{max}-q_{0}|$ is found, one can find using the Bragg condition the maximal reflectivity bandwidth in the wavelength or angle domain for the specified value of $r_c$, and then the maximal integral reflectivity. 

The maximum inverse period range $\Delta q_{max}$ can be estimated by a direct integration in the equations (\ref{fz2}) or (\ref{fz6}). Neglecting for simplicity the material dispersion we have
\begin{equation}
\Delta q_{max}\leqslant-\pi \bar{q}\frac{|B|^{2}}{\gamma}\frac{1}{\mu'}\int\limits_{s}^{1}\frac{du}{u}\frac{1}{\ln\left[\frac{1-u}{1+u}\right]},
\label{ch3eq2g4}
\end{equation}
where $\bar{q}$ is some average value of $q$ and the lower integration boundary is defined as $s=r_{c}/(1-r_{c})$.

Using the formula (\ref{ch3eq2g4}) as well as the Bragg condition (\ref{fn}) (taken at $B=0$), both wavelength and angle bandwidths can be expressed as
\begin{eqnarray}
\Delta\lambda&\leqslant&-\pi\bar{\lambda}\frac{|B|^{2}}{\gamma}\frac{1}{\mu'}\int\limits_{s}^{1}\frac{du}{u}\frac{1}{\ln\left[\frac{1-u}{1+u}\right]},\label{ch3eq2g6}\\
\Delta\theta&\leqslant&-\pi\bar{\theta}\frac{|B|^{2}}{\gamma}\frac{1}{\mu'}\int\limits_{s}^{1}\frac{du}{u}\frac{1}{\ln\left[\frac{1-u}{1+u}\right]},\label{ch3eq2g7}
\end{eqnarray}
respectively, where parameters $B$, $\mu$ and $\gamma$ should be taken in some middle point $\bar{\lambda}$ or $\bar{\theta}$. 

Let us consider the case when $r_c\ll1$ and, therefore, $s\ll1$. Then the common integral in (\ref{ch3eq2g6}) and (\ref{ch3eq2g7}) can be evaluated as
\begin{equation}
I=\int\limits_{s}^{1}\frac{du}{u}\frac{1}{\ln\left[\frac{1-u}{1+u}\right]}\approx\frac{1}{2}\left[\frac{1}{s}-1-\frac{1}{\sqrt{3}}\left(\frac{\pi}{6}-\mbox{atan}\frac{s}{\sqrt{3}}\right)
\right],
\label{ch3eq2g8}
\end{equation}
which diverges when $s\to0$. This is not surprising as the bandwidth is expected to become progressively wider when the reflectivity decreases. If $r_c$ were zero, the bandwidth would be  infinite.

Let us multiply (\ref{ch3eq2g8}) by $r_c$ and then substitute the result into \eqref{ch3eq2g6} or \eqref{ch3eq2g7}. The outcome is an estimate of the maximal integral reflectivity over a wavelength or grazing angle range
\begin{align}
r_{int,\lambda}&\leqslant\frac{\pi}{2}\bar{\lambda}\frac{|B|^{2}}{\gamma\mu'}\left[1-r_c\left(2+\frac{\pi}{2\sqrt{3}}\right)+\frac{r_c}{\sqrt{3}}\operatorname{atan}\frac{r_c}{\sqrt{3}(1-r_c)}\right],\label{ch3eq2g9}\\
r_{int,\theta}&\leqslant\frac{\pi}{2}\bar{\theta}\frac{|B|^{2}}{\gamma\mu'}\left[1-r_c\left(2+\frac{\pi}{2\sqrt{3}}\right)+\frac{r_c}{\sqrt{3}}\operatorname{atan}\frac{r_c}{\sqrt{3}(1-r_c)}\right].\label{ch3eq2g9a}
\end{align}
The average wavelength or grazing angle is assumed to be equal to $\bar{\lambda}=\sqrt{\lambda_1\lambda_2}$ or $\bar{\theta}=\sqrt{\theta_1\theta_2}$, respectively, where $\lambda_1$ and $\lambda_2$ or $\theta_1$ and $\theta_2$ are the wavelengths or grazing angles, corresponding to the boundaries of the respective reflectivity bands. 

From the formulas (\ref{ch3eq2g9}) and (\ref{ch3eq2g9a}) it follows that, when $r_c$ decreases, the integral reflectivity $r_{int}$ increases reaching the maximum at $r_c=0$. Thus the limiting maximal value of the integral reflectivity can be obtained by substituting $r_c=0$ into \eqref{ch3eq2g9} or \eqref{ch3eq2g9a}
\begin{align}
r_{int,\lambda}^{lim}&\leqslant\frac{\pi}{2}\bar{\lambda}\frac{|B|^{2}}{\gamma\mu'},\label{ch3eq2g10}\\
r_{int,\theta}^{lim}&\leqslant\frac{\pi}{2}\bar{\theta}\frac{|B|^{2}}{\gamma\mu'}.\label{ch3eq2g10a}
\end{align}
The maximal values of the integral reflectivities calculated using (\ref{ch3eq2g10}) are shown in Table \ref{t1} in the last column, while the respective average values of the wavelength or grazing angle in the previous column. From Table \ref{t1} it can be seen that the integral reflectivities of real structures are 10--40\% smaller than the estimated maximal values, especially in the soft X-ray range.

\begin{table}[tb]
\caption{The integral reflectivities (in nm) for the numerical experiments, whose results are shown in Fig.\ref{f3}--\ref{f4} (columns 2--5). The column 6 shows the typical integral reflectivities of the periodic multilayer structures optimized for the respective wavelength ranges. The column 7 shows the average wavelength used and the column 8 -- the upper limits (in nm) for these average wavelengths calculated using the formula (\ref{ch3eq2g10}) with $\beta=0.4$.} 
\label{t1}
\begin{tabular}{cccccccc}
\toprule
\textbf{\thead{Figure/\\Panel}}&\textbf{a}&\textbf{b}&\textbf{c}&\textbf{d}&\textbf{\thead{Periodic\\structure}}&$\bar{\mathbf{\lambda}}$, \textbf{nm}&\thead{$\mathbf{r_{int}^{lim}}$,\\ \textbf{nm}}\\
\midrule
\ref{f3}&0.00176&0.00329&0.00219&0,00306&0.001--0.015&0.08&0.00431\\
\ref{f4}&0.02073&0.01623&0.01648&0.01664&0.015--0.017&4.5&0.0316\\
\bottomrule
\end{tabular}
\end{table}

Formulas (\ref{ch3eq2g10}) and (\ref{ch3eq2g10a}) can be also used to obtain the optimal value of $\beta$, at which the maximal integral reflectivity is reached. Noting that (\ref{ch3eq2g10}) and (\ref{ch3eq2g10a}) depend on $\beta$ only through a factor
$$
m=|B|^{2}/\gamma=\frac{|\varepsilon_{1}-\varepsilon_{2}|^{2}}{4\pi^{2}}\frac{\sin^{2}\pi\beta}{\operatorname{Im}(\varepsilon_{2})+\beta(\varepsilon_{1}-\varepsilon_{2})}.
$$
Differentiating the latter expression with respect of $\beta$ and equaling the result to zero we can obtain the following equation for $\beta_{max}$ 
\begin{equation}
\tan\pi\beta_{max}=2\pi(\beta_{max}+g),
\label{ch3eq2g11}
\end{equation}
where $g=\operatorname{Im}\varepsilon_{2}/\operatorname{Im}(\varepsilon_{1}-\varepsilon_{2})$. When one material in a multilayer structure has the absorption much large than the second material and therefore $g\ll1$, the solution of \eqref{ch3eq2g11} is $\beta_{max}\approx0.37$. 

Let us note that while (\ref{ch3eq2g11}) was obtained for the reflectivities having table-like shapes, it will be approximately valid for any shape resembling the table-like profile. 

\section{Conclusion}
\pst
An analytical theory of the graded multilayer coatings based on the coupled wave approximation was developed. A number of exact solutions of the coupled wave equations for different dependencies of the inverse period on the depth were found. A series of numerical experiments demonstrated a good agreement between the graded multilayer reflectivities calculated using the obtained exact solutions with the calculations by the exact matrix method. The coupled wave amplitudes were then approximated by WKB asymptotics and approximate formulas for the multilayer reflectivity were obtained. The numerical experiments then showed that the use of the WKB asymptotics leads to the reflectivities that are in a satisfactory agreement with the matrix method calculations especially within the resonant band. It was also revealed that a significant widening of the reflectivity band as compared to the periodical multilayers is achievable. 

As the next step differential equations for solution of the inverse problem for graded multilayers in the wavelength or angle domain were derived based on the developed analytical theory. A number of numerical experiments, which demonstrated the suitability of this approximate method for the real world inverse problem solution in different wavelength ranges, was conducted. They demonstrated a good agreement of the obtained inversions with the target reflectivities. The general problem of the integral reflectivity maximization was then discussed and a general second order differential equation for its solution was derived from the variational principle. Finally, a formula, which estimates the upper bound of the achievable integral reflectivity, was obtained and compared to the results of the numerical experiments.

To summarize, the developed analytical theory of graded multilayers can be used to design non-periodic X-ray multilayers in a wide range of wavelengths (from 0.1 to 10 nm) and grazing angles. The inversions obtained with it can be further rectified using standard numerical optimization methods, for instance, the fastest descent method. This means that the presented theory can potentially facilitate the development of the broadband X-ray reflective optics for such applications as the \enquote{water window} and \enquote{carbon window} microscopes as well as for the X-ray telescopes working in the hard part of the X-ray spectrum.

\section*{Acknowledgments}
\pst
The authors are thankful to Prof. A.V. Popov and Dr. I.V. Kozhevnikov for fruitful discussions.

\end{document}